%% file: iclr2027_conference.tex
\documentclass{article} % For LaTeX2e
\usepackage{iclr2027_conference,times}

\input{math_commands.tex}

\usepackage{hyperref}
\usepackage{url}
\usepackage{enumerate}
\usepackage{enumitem}
\usepackage{amsthm}
\usepackage{graphicx}
\usepackage{booktabs}
\usepackage{multirow}

\usepackage{amssymb} 

\usepackage{algorithm}
\usepackage{algpseudocode}

\usepackage{array}
\usepackage{tabularx}
\usepackage{makecell}

\usepackage[dvipsnames]{xcolor}
\usepackage{hyperref}
\usepackage{url}

\hypersetup{
  colorlinks=true,
  allcolors=MidnightBlue
}

\usepackage{marvosym}

\title{Plan-to-Synthesis: Cross-City Human Mobility Generation via Semantic Latent Flow Matching}

\iclrfinalcopy
\author{Zhoufu Wang$^{1,2}$\thanks{Equal contribution. \textsuperscript{\Letter}Corresponding author.} \quad 
  \textbf{Baoshen Guo}$^{1}$\textsuperscript{\Letter}\footnotemark[1] \quad
  Zhiqing Hong$^{4}$ \quad
  Junyi Li$^{1}$ \quad
  Kailai Sun$^{1}$ \quad \\
  \textbf{Heye Huang$^{3}$ \quad
  Alok Prakash$^{1}$ \quad 
  Shenhao Wang$^{5}$ \quad
  Jinhua Zhao$^{6}$}\\
  $^{1}$Singapore-MIT Alliance for Research and Technology Centre, MIT\\
  $^{2}$Nanyang Technological University
  \quad $^{3}$Korea Advanced Institute of Science \& Technology
  \\
  $^{4}$Hong Kong University of Science \& Technology (Guangzhou)
  \quad$^{5}$University of Florida\\
  $^{6}$Massachusetts Institute of Technology\\
}

\begin{document}

\maketitle
\lhead{Preprint} 

\begin{abstract}

Human mobility generation aims to synthesize realistic point-of-interest (POI) visitation trajectories and has become an important tool for travel behavior modeling, transportation management, and urban planning. 
Existing diffusion-based methods achieve high fidelity but require per-city generation, given the inherent heterogeneity of geospatial locations and POI categories, while large language model-based methods generalize across cities but remain too costly at scale, especially for long-horizon trajectory generation. 
To address this, we propose \textit{SeMoFlow}, a \textbf{Se}mantic human \textbf{Mo}bility generation framework based on latent \textbf{Flow} matching. 
We first encode heterogeneous POIs from different cities into a shared cross-city representation space via hierarchical Semantic IDs, where shared prefixes capture transferable semantics, and successive codes progressively refine the representation toward individual POIs.
Building on the semantic IDs, \textit{SeMoFlow} follows a plan-to-synthesis hierarchical generation paradigm, in which an autoregressive planner generates coarse-grained semantic and recurrence patterns, and a flow matching realizer synthesizes
fine-grained suffix latents. The generated latents
are subsequently decoded and grounded
to concrete POIs.
Extensive experiments on large-scale multi-city datasets show that SeMoFlow achieves higher trajectory fidelity than existing baselines, preserves city-specific mobility motifs, and supports both joint multi-city generation and effective cross-city transfer. 
The code is available at \href{https://github.com/urban-mobility-generation/SeMoFlow}{https://github.com/urban-mobility-generation/SeMoFlow}.
\end{abstract}

\input{01-Intro}
\input{02-Related-Work}

\input{03-Problem}

\input{04-Design}

\input{05-Evaluation}

\input{06-Conclusion}

\section*{AI Use Statement}
Generative AI tools were used to assist with language editing, consistency
checks, and grammar correction. The authors reviewed and verified all
AI-assisted text and remain responsible for the scientific claims,
experimental design, results, citations, and final manuscript. 

\section*{Reproducibility Statement}
The anonymized source code, including training and evaluation scripts, is
provided in the supplementary material. The model architecture, training
objective, hyperparameters, network
configurations, and optimization settings are listed in Appendix Sec.~\ref{app:setup}. We also release the generated synthetic trajectory data for
each city in our experiments in the supplementary material.

\bibliography{iclr2027_conference}
\bibliographystyle{iclr2027_conference}

\newpage

\appendix
\input{07-Appendix}

\end{document}

%% file: math_commands.tex
\usepackage{amsmath,amsfonts,bm}

\def\eqref#1{equation~\ref{#1}}
\def\1{\bm{1}}

\DeclareMathAlphabet{\mathsfit}{\encodingdefault}{\sfdefault}{m}{sl}
\SetMathAlphabet{\mathsfit}{bold}{\encodingdefault}{\sfdefault}{bx}{n}

%% file: 01-Intro.tex
\section{Introduction}\label{sec:intro}

% background and problem

Human mobility trajectories, which record citizens' daily activities as sequences of visited points of interest (POIs), play a crucial role in smart city applications such as travel demand estimation, urban planning, and socioeconomic analysis~\citep{guo2025language}. However, large-scale real-world mobility data are rarely publicly available due to the cost of collection and the exposure of sensitive locations and individual behavioral patterns~\citep{yabe2024yjmob100k}.
Generating synthetic mobility trajectories has therefore emerged as an important research direction, aiming to produce large-scale POI visit sequences that align with real-world mobility distributions for downstream utility.

% related work and limitations

Recently, several works have leveraged generative AI for human activity trajectory synthesis, which can be divided into semantic-agnostic and semantic-aware methods. 
Semantic-agnostic methods represent mobility as a geometric process over GPS coordinates~\citep{zhu2023difftraj,controltraj2024} or road segments~\citep{wei2024diff,guo2026leveraging,wongso2026trajdlm}, which reproduce movement dynamics with diffusion models~\citep{guo2026leveraging} or flow matching models~\citep{li2026trajflow} without describing what each visit means.
Semantic-aware generative methods target POI sequences, where each visit carries an explicit activity meaning. 
For example, some works encode visit sequences into continuous latents and apply diffusion or flow matching over them~\citep{xu2026synhat,song2024controllable}, while others perform discrete diffusion directly over POI token sequences~\citep{xu2026mobidiff}. 
However, these works can hardly achieve joint multi-city training and generation, or transfer to unseen cities, given the heterogeneity of POI ID vocabularies and geospatial partitions across cities.
In addition, 
existing LLM-based mobility generation methods~\citep{wang2026ellmob,wang2024llmob,liu2026allcities} condition on user profiles, intentions, or events, and leverage the transferability of language semantics, showing potential for cross-city generation, but incur prohibitive cost when generating large-scale, long-horizon human activities.

% challenges, intuition
In this paper, we target semantic-aware human activity trajectory generation across cities, which remains difficult at scale due to: 
(i) \textbf{ Cross-city semantic heterogeneity:} POI semantics are heterogeneous across cities. Identifiers from different cities are disjoint and carry no inherent correspondence, while collapsing POIs into coarse categories makes them comparable but discards the individual identity that a generated visit must eventually resolve to. 
(ii) \textbf{Diverse recurrence structure:} Human mobility exhibits recurring visits to familiar locations, often following regular daily or weekly routines~\citep{schlapfer2021universal}. However, recurrence patterns vary across locations and temporal scales, making them difficult to capture through direct trajectory generation.

% our key idea, design, and detailed contributions 

To address these challenges, we propose \textit{SeMoFlow}, a semantic human mobility generation framework. 
Specifically, to resolve cross-city semantic heterogeneity, we encode POIs from multiple cities into a unified representation space through hierarchical Semantic IDs, where shared prefixes group functionally similar POIs
across cities, while successive codes progressively refine
their representations toward individual POIs. 
To capture diverse recurrence patterns, we introduce a plan-to-synthesis framework. 
An autoregressive planner models recurring visits through
Semantic ID prefixes, while a flow-matching realizer
synthesizes fine-grained suffix latents conditioned on the
planned prefixes.
In summary, our key contributions are as follows.
\begin{itemize}[leftmargin=*]
\item We are the first to leverage Semantic IDs to encode POIs from heterogeneous cities into a unified hierarchical representation space for cross-city mobility generation, upon which we build SeMoFlow, a semantic-aware human activity generation framework with latent flow matching.
\item We propose a hierarchical plan-to-synthesis generation paradigm aligned with the Semantic ID hierarchy, where an autoregressive planner plans over code prefixes and a
latent flow-matching realizer synthesizes fine-grained
suffix representations for subsequent decoding and
POI grounding.
\item Experiments on large-scale datasets covering multiple cities show that SeMoFlow surpasses existing studies in generation fidelity and mobility motif preservation. We further analyze the interpretability of the learned semantic codes and the cross-city transferability of the plan-to-synthesis design. Code and synthetic data are provided in the supplementary material.

\end{itemize}

%% file: 02-Related-Work.tex
\section{Related Works}\label{sec:related-works}

\subsection{Human Mobility Generation}

Human mobility generation is essential for smart cities~\citep{zhu2023difftraj,guo2026leveraging,li2026trajflow}. 
Recent advances in diffusion models~\citep{ho2020denoising}, flow matching~\citep{lipman2023flowmatching,li2026trajflow}, and large language models~\citep{wang2024llmob,wang2026ellmob} have been applied to mobility trajectory synthesis. 
Existing research can be divided into semantic-agnostic and semantic-aware generation. 
Semantic-agnostic methods treat mobility as movement through physical space, generating GPS coordinates or road segments without describing what a visit means. DiffTraj~\citep{zhu2023difftraj} and ControlTraj~\citep{controltraj2024} generate GPS trajectories under spatio-temporal and road-topology conditioning, while TrajFlow~\citep{li2026trajflow} adopts flow matching at nationwide scale. 
% These methods reproduce movement dynamics with high fidelity, but their state space is tied to a city-specific coordinate system or road network. 
Semantic-aware methods instead generate POI sequences in which each visit corresponds to an activity. 
GeoGen~\citep{xu2026geogen} and SynHAT~\citep{xu2026synhat} realize semantic activities through coarse-to-fine stages.
Traveller reproduces recurrent visits with an anchor-conditioned discrete diffusion realizer~\citep{luo2026traveller}.  
However, their activity semantics are defined within a single city and provide no correspondence across urban catalogs.
LLM-based mobility generation~\citep{wang2024llmob,wang2026ellmob,liu2026allcities} further condition on personal patterns or events at a higher computational cost, especially for long-horizon (e.g., weekly, monthly) generation. 

\subsection{Cross-city Human Mobility Modeling and Applications}
Cross-city modeling is essential for building unified foundation models of urban mobility, yet most existing efforts target spatio-temporal representation and prediction. ME-POIs~\citep{siampou2026mepois} learns transferable place-function representations by augmenting text-derived POI embeddings with large-scale mobility signals, and TransferTraj~\citep{wei2025transfertraj} studies region and task transferability for vehicle trajectories. COLA~\citep{wang2024cola} separates city-shared and city-private Transformer components to transfer mobility knowledge across cities. 
For cross-city generation, \cite{liu2026allcities} plan trips with an LLM and generate trajectories with diffusion over a shared spatial representation. In this work, we study cross-city mobility generation at the POI level, which remains challenging because POIs carry heterogeneous semantics across disjoint city vocabularies, and generation must decode a concrete identity from a large discrete space. 

\subsection{Semantic IDs for Generative Modeling}
In recommendation systems, semantic IDs~\citep{rajput2023tiger,ju2025generative} encode the semantic content of an item into a tuple of coarse-to-fine discrete codes through residual quantization, rather than assigning it an arbitrary index.
TIGER~\citep{rajput2023tiger} first replaces atomic item identifiers with such codes and generates them autoregressively, turning retrieval into generation over a compact vocabulary. Within a code tuple, a shared prefix groups semantically similar items, while the full tuple identifies an individual item~\citep{ju2025generative,wei2025cofirec}.
For mobility studies, semantic IDs have been introduced for next-POI recommendation tasks~\citep{wang2025gnprsid,chen2025enhancing}, which usually encode POIs into code sequences from semantic and collaborative features to alleviate the sparsity of atomic identifiers. 
However, 
from the POI-level trajectory generative perspective, ours is the first work to leverage semantic IDs to place POIs from different cities in one representation space and to generate fine-grained trajectories over it.

%% file: 03-Problem.tex
\section{Preliminary}\label{sec:problem}

\subsection{Problem Formulation}

\noindent \textit{Human Mobility Trajectory:} A human mobility trajectory records the POI visiting activities of an anonymous individual, denoted as $\tau = (s_1, s_2, \cdots, s_{L})$, where $s_i = (t_i, p_i)$ indicates a visit to POI $p_i$ at time $t_i$ and $L$ is the trajectory length. Each POI $p \in \mathcal{P}_c$ belongs to the catalog $\mathcal{P}_c$ of city $c$, and is associated with its geographic location, address, name, and functional categories.

\noindent \textit{Cross-city Human Mobility Generation:} Given a set of cities $\mathcal{C}$ and the observed trajectories $\mathcal{T}_c = \{\tau_1, \tau_2, \cdots, \tau_{N_c}\}$ of each city $c \in \mathcal{C}$, our goal is to learn a unified generative model $\mathcal{F}$ over all cities. Conditioned on a target city $c$, $\mathcal{F}$ synthesizes trajectories $\hat{\mathcal{T}}_c$ that are grounded in the catalog $\mathcal{P}_c$, match the mobility distribution of $\mathcal{T}_c$, and preserve its semantic activity patterns.

\subsection{Mobility Semantics: Coarse-to-fine and Recurrence Patterns}

\begin{figure}[t]
    \centering
    \includegraphics[width=0.9\linewidth]{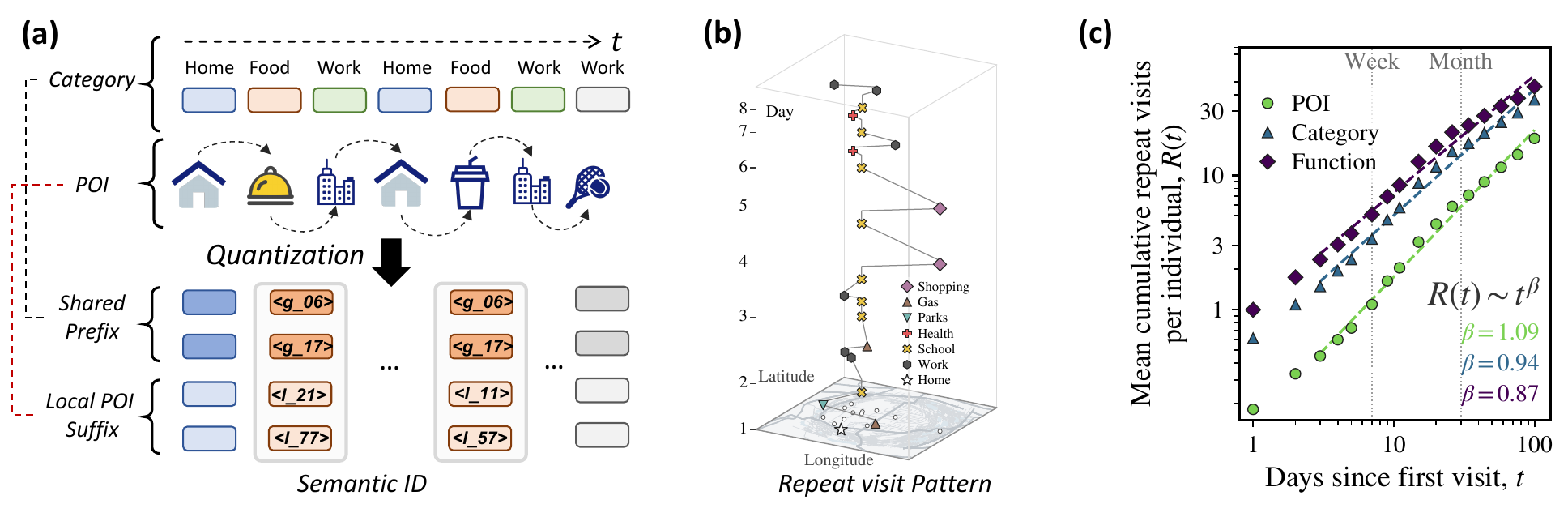}
    \caption{(a) A trajectory admits a coarse function or category sequence and a fine POI sequence. Residual quantization aligns these two levels with Semantic IDs, where a shared prefix encodes the coarse-grained semantics (e.g., categories) and a local suffix resolves the concrete POI (e.g., two \textit{food visits} share the prefix but differ in their suffix). (b) An individual trajectory over space and time, where a set of POIs is visited repeatedly. (c) Mean cumulative repeat visits per individual, and recurrence is consistently stronger at the function and category levels than at the POI level.}
    \label{fig:motivation}
\end{figure}

\textbf{Hierarchy of Human Mobility Semantics}: Human mobility carries semantics at multiple granularities. 
As illustrated in Fig.~\ref{fig:motivation}(a), a visit can be described coarsely by the activity it serves, such as dining or shopping, and finely by the specific place at which the activity occurs. The two levels play different roles in generation. Coarse activity semantics are comparable across cities, since every city hosts restaurants and workplaces, whereas the fine level is bound to a city-specific catalog. 

\textbf{Mobility Recurrence Patterns}: 
Fig.~\ref{fig:motivation}(b) and (c) show that human mobility also exhibits intrinsic recurrence. Individual movement is constrained by regular spatial preferences and preferential return, and daily movement falls into a small number of stable motifs~\citep{gonzalez2008mobility,schneider2013unravelling}. Recurrence arises at both granularities, as a person may commute to the same office every weekday while having lunch at a different restaurant each time.

\subsection{Latent Flow Matching for Discrete Sequences}

Flow matching learns a continuous-time vector field that
transports samples between a noise distribution and the
data distribution along a probability path~\citep{lipman2023flowmatching}.
To apply flow matching to discrete sequences, let a discrete
sequence $\mathbf{s}=[s_1,\ldots,s_L]\in\mathcal{V}^L$ be
mapped to a continuous latent representation
$\mathbf{x}\in\mathbb{R}^{L\times d}$, and let
$\boldsymbol{\epsilon}\sim\mathcal{N}(\mathbf{0},\mathbf{I})$.
Following Rectified Flow~\citep{liu2022flow}, we define
$
\mathbf{z}_{\eta}
=
(1-\eta)\mathbf{x}
+
\eta\boldsymbol{\epsilon},
\quad
\mathbf{v}^{\star}
=
\boldsymbol{\epsilon}-\mathbf{x},
$
where $\eta=0$ corresponds to the data endpoint and
$\eta=1$ to the noise endpoint. Under the
$\mathbf{x}$-prediction parameterization, a neural network
$\mathbf{x}_{\theta}(\mathbf{z}_{\eta},\eta)$ predicts the
clean latent, yielding the velocity estimate
$
\mathbf{v}_{\theta}(\mathbf{z}_{\eta},\eta)
=
\frac{
\mathbf{z}_{\eta}
-
\mathbf{x}_{\theta}(\mathbf{z}_{\eta},\eta)
}{
\bar{\eta}
},
\quad
\bar{\eta}=\max(\eta,\eta_{\min}),
$
where $\eta_{\min}>0$ is a small constant for numerical
stability. The model is trained by matching this velocity
to $\mathbf{v}^{\star}$. At inference time, the learned
flow is integrated from $\eta=1$ to $\eta=0$ to transform
Gaussian noise into a latent sequence.

%% file: 04-Design.tex
\section{Model Design}\label{sec:model}

\begin{figure}[h]
    \centering
    \includegraphics[width=0.95\linewidth]{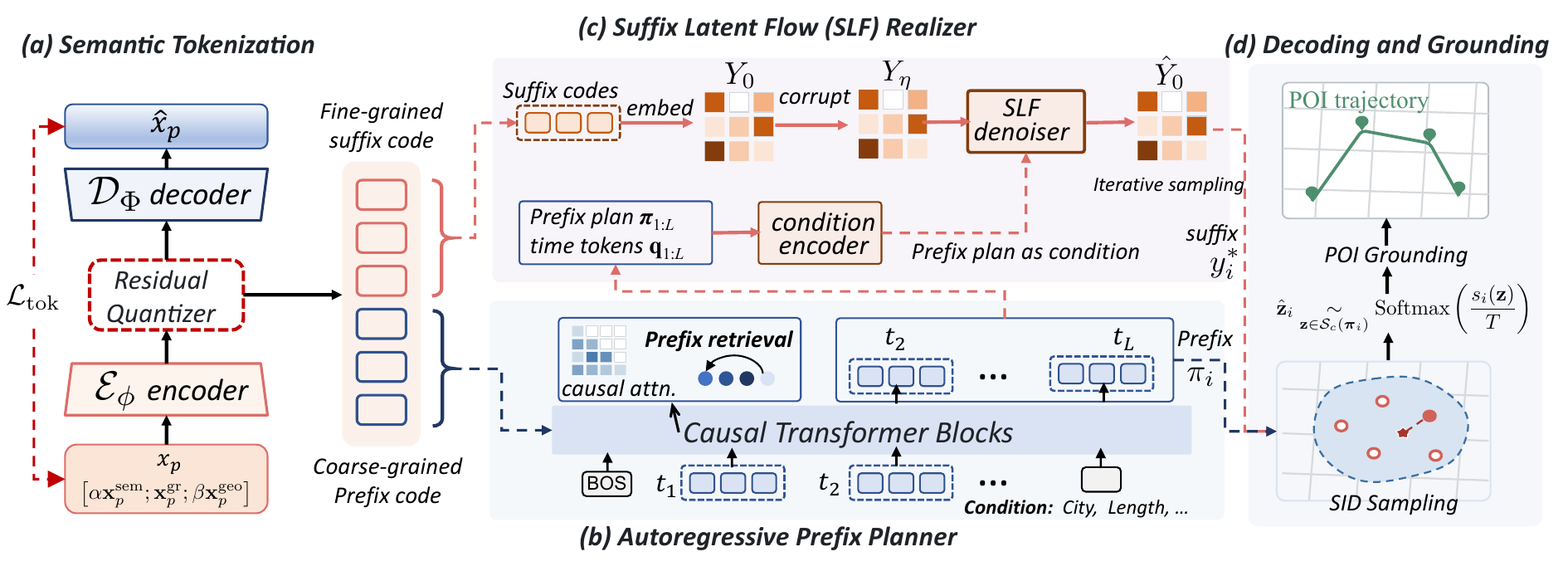}
\caption{
\textbf{Overview of \textit{SeMoFlow}.}
(a) A semantic tokenizer encodes POI features into
hierarchical Semantic IDs.
(b) An autoregressive planner generates time gaps
and SID prefixes to form a coarse-grained trajectory plan.
(c) A conditional latent flow realizer learns to denoise
corrupted suffix latents and generates suffix
representations.
(d) The generated suffix latents are decoded into
full SIDs under the planned prefixes and subsequently
grounded to concrete POIs.
}
    \label{fig:framework}
\end{figure}

Fig.~\ref{fig:framework} presents an overview of the \textit{SeMoFlow} framework.
We first represent each POI by a Semantic ID (Sec.~\ref{sec:sid}), a residual-quantized code whose prefix
levels describe an activity in an area and whose suffix levels refine the representation toward individual POIs, to represent cross-city POIs in a unified latent space.
Then, in Sec.~\ref{sec:planner} and Sec.~\ref{sec:realizer}, we introduce our coarse-to-fine generative design, in which an autoregressive planner generates the SID prefixes
visit by visit, while a conditional latent flow matching
realizer generates suffix representations.
The generated latents are then decoded into full SIDs and grounded to POIs in the target city.

\subsection{Hierarchical semantic tokenization for POI representation}
\label{sec:sid}

Human trajectories can be viewed as a physical language, where POI visits serve as tokens of human activities. However, city-specific POI identifiers lack transferable semantics, hindering cross-city trajectory generation. Inspired by Semantic IDs (SIDs) in generative recommendation~\citep{rajput2023tiger,wang2025gnprsid}, we encode POI semantics and spatial attributes into hierarchical discrete codes via residual quantization~\citep{lee2022autoregressive,wei2025cofirec}. By jointly learning the tokenizer across cities, we establish a shared semantic code space in which common prefixes capture transferable activity semantics, while successive codes refine toward individual POI identities.

\paragraph{Input features.}
For each POI $p$, we construct its representation from three sources: (i) semantic features derived from its name, brand, and category; (ii) graph embeddings learned from a POI transition graph; and (iii) city-relative geographic features based on offsets from the city centroid.
After PCA-based dimensionality reduction and feature standardization, we have
$
    \mathbf{x}_p =
    \left[
    \alpha\mathbf{x}^{\mathrm{sem}}_p;
    \mathbf{x}^{\mathrm{gr}}_p;
    \beta\mathbf{x}^{\mathrm{geo}}_p
    \right]
$, where $\mathbf{x}^{\mathrm{sem}}_p$, $\mathbf{x}^{\mathrm{gr}}_p$, and $\mathbf{x}^{\mathrm{geo}}_p$ denote the embedded textual semantic, graph, and geographic features, respectively, and $\alpha$ and $\beta$ control their relative contributions.

\begin{figure}[h]
    \centering
    \includegraphics[width=0.9\linewidth]{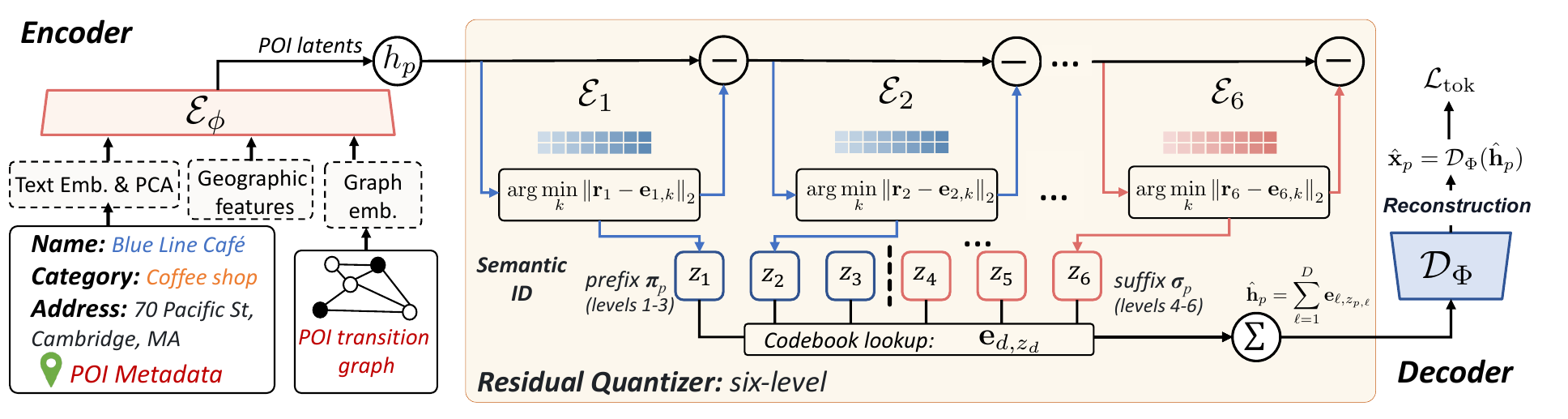}
\caption{Hierarchical SID tokenization of POIs.
Shared prefix codes capture transferable activity
semantics, while successive suffix codes refine
the representation toward individual POI identities.}
    \label{fig:sid}
\end{figure}

\paragraph{Hierarchical semantic tokenization via residual quantization.}
Given the POI representation $\mathbf{x}_p$, we employ residual quantization to learn hierarchical Semantic IDs.
Specifically, an encoder $\mathcal{E}_{\phi}$ maps $\mathbf{x}_p$ into a latent representation $\mathbf{h}_p=\mathcal{E}_{\phi}(\mathbf{x}_p)$, which is progressively quantized using $D$ codebooks $\mathcal{E}_{\ell}=\{\mathbf{e}_{\ell,k}\}_{k=1}^{K}$.
Starting from $\mathbf{r}_{p,1}=\mathbf{h}_p$, each level selects the nearest code $z_{p,\ell}=\arg\min_{k\in[K]}\|\mathbf{r}_{p,\ell}-\mathbf{e}_{\ell,k}\|_2^2$ and updates the residual as $\mathbf{r}_{p,\ell+1}=\mathbf{r}_{p,\ell}-\mathbf{e}_{\ell,z_{p,\ell}}$.
The resulting SID $\mathbf{z}_p=(z_{p,1},\ldots,z_{p,D})\in[K]^D$ represents the POI hierarchically, with its quantized latent given by $\hat{\mathbf{h}}_p=\sum_{\ell=1}^{D}\mathbf{e}_{\ell,z_{p,\ell}}$.
Following this coarse-to-fine hierarchy, we partition each SID into:
\begin{equation}
\mathbf{z}_p=
\bigl(
\underbrace{z_{p,1},\ldots,z_{p,P}}_{\text{prefix }\boldsymbol{\pi}_p}
\;\big|\;
\underbrace{z_{p,P+1},\ldots,z_{p,D}}_{\text{suffix }\boldsymbol{\sigma}_p}
\bigr).
\label{eq:split}
\end{equation}
The prefix captures coarse-grained activity characteristics shared across POIs, providing transferable semantics for modeling recurring activity patterns, while the suffix progressively refines the representation toward individual POI identities.

\paragraph{Tokenizer training.}
We train the tokenizer to reconstruct the POI features from the quantized latent, with a commitment loss that aligns the encoder outputs with the selected code vectors:
\begin{equation}
\mathcal{L}_{\mathrm{tok}}
=
\left\|\mathbf{x}_p-\mathcal{D}_{\Phi}(\hat{\mathbf{h}}_p)\right\|_2^2
+
\lambda_{\mathrm{com}}
\sum_{\ell=1}^{D}
\left\|\mathbf{r}_{p,\ell}
-\operatorname{sg}[\mathbf{e}_{\ell,z_{p,\ell}}]\right\|_2^2,
\label{eq:tok_loss}
\end{equation}
where $\operatorname{sg}[\cdot]$ denotes the stop-gradient operator. We also follow standard VQ-VAE training~\citep{oord2017neural} with balanced assignments~\citep{cuturi2013sinkhorn} to improve codebook utilization.

\subsection{Recurrence-aware Autoregressive Prefix Planner}
\label{sec:planner}

After POI semantic tokenization, 
as shown in Fig.~\ref{fig:framework}, 
we introduce a recurrence-aware autoregressive planner to generate coarse-grained trajectories over SID prefixes.

\paragraph{Autoregressive prefix planning.}
Given a target city $c$, the planner first samples the trajectory length $L$, and then autoregressively generates a coarse-grained trajectory $\{(g_i,\boldsymbol{\pi}_i)\}_{i=1}^{L}$ through
$
p_\theta(L,g_{1:L},\boldsymbol{\pi}_{1:L}\mid c)
=p_\theta(L\mid c)\prod_{i=1}^{L}
p_\theta(g_i\mid\mathbf{h}_i)
p_\theta(\boldsymbol{\pi}_i\mid\tilde{\mathbf{h}}_i,\mathcal{M}_i),
\label{eq:planner_likelihood}
$ where 
$g_i$ denotes the discretized time gap, $\boldsymbol{\pi}_i$ is the SID prefix of the $i$-th visit,
$\mathbf{h}_i$ encodes preceding visits, $\tilde{\mathbf{h}}_i$ incorporates the sampled gap and updated time, and $\mathcal{M}_i$ denotes the history of visited prefixes.

To plan successive visits, we use causal Transformer blocks to encode the preceding visits, represented by their prefix codes, time gaps, together with the trajectory length and city context, into \(\mathbf{h}_i\). The planner samples the next gap \(g_i\) and incorporates the updated time into \(\tilde{\mathbf{h}}_i\) to predict the next prefix \(\boldsymbol{\pi}_i\). 
The updated visit time is
further discretized into a weekly temporal token $q_i$.

\paragraph{Explore--Retrieve prefix generation.}
To capture the exploration and recurrence patterns of human activities, we introduce two complementary mechanisms for prefix generation: (i) \textit{Explore}, which generates a prefix from the shared semantic vocabulary, and (ii) \textit{Retrieve}, which reuses a previously visited prefix. Their distributions are combined through a learnable gate $r_i\in[0,1]$ to adaptively balance generating activities and revisiting the existing ones:
\begin{equation}
p_\theta(\boldsymbol{\pi}_i\mid\tilde{\mathbf{h}}_i,\mathcal{M}_i)
=
(1-r_i)p_i^{\mathrm{exp}}(\boldsymbol{\pi}_i)
+r_i p_i^{\mathrm{ret}}(\boldsymbol{\pi}_i).
\label{eq:router}
\end{equation}

\textit{Explore.}
Conditioned on $\tilde{\mathbf{h}}_i$, a GRU-based~\citep{chung2014empirical} decoder autoregressively generates the $P$ prefix codes, where each code is predicted from the preceding levels. The output vocabulary is constrained by the target city's POI catalog, ensuring that every generated prefix admits a valid POI completion.

\textit{Retrieve.}
Instead of generating a prefix from scratch, the retrieve branch selects a previously visited prefix from the historical memory $\mathcal{M}_i$. We employ three complementary retrieval experts: (i) a neural expert that matches the current state to historical occurrences, (ii) a frequency expert that favors frequently visited prefixes, and (iii) a last-visit expert that favors the immediately preceding prefix. Together, these experts capture diverse recurrence patterns in human mobility.

\paragraph{Training objective.}
We train the planner by teacher forcing to maximize the likelihood of the observed trajectory length, time gaps, and SID prefixes:
\begin{equation}
\mathcal{L}_{\mathrm{plan}}
= -\log p_\theta(L\mid c)
-\sum_{i=1}^{L}\left[
\log p_\theta(g_i\mid\mathbf{h}_i)
+\log p_\theta(\boldsymbol{\pi}_i\mid\tilde{\mathbf{h}}_i,\mathcal{M}_i)
\right].
\label{eq:planner_loss}
\end{equation}
The explore-retrieve decision is learned implicitly through the mixture likelihood in Eq.~(\ref{eq:router}).

\subsection{Conditional Latent Flow Matching-based Realizer}
\label{sec:realizer}

Given the coarse-grained plan $(\boldsymbol{\pi}_{1:L},q_{1:L})$, 
we further design a suffix latent flow (SLF) matching-based realizer to synthesize fine-grained suffix representations to resolve individual POI visits. Rather than generating discrete suffix codes independently, we employ conditional latent flow matching to jointly model the suffixes of all occurrences in the continuous semantic space learned by the tokenizer.

\paragraph{Coarse-grained plan encoding.}
To guide fine-grained POI generation, we encode the planned SID prefixes and temporal tokens into a condition sequence $\mathbf{C}\in\mathbb{R}^{L\times d}$ using a bidirectional Transformer $F_{\mathrm{cond}}$. Specifically, each occurrence is represented by the sum of its prefix code embeddings, temporal embedding, positional embedding, and city embedding. The bidirectional attention captures dependencies across the entire planned trajectory, providing global semantic and temporal context for subsequent suffix generation.

\paragraph{Conditional latent flow matching.}
Given the planned prefixes, we represent each POI suffix in the tokenizer's continuous latent space as
$\mathbf{y}_i=\mathrm{Norm}(\sum_{\ell=P+1}^{D}\mathbf{e}_{\ell,z_{i,\ell}})$,
where $\mathrm{Norm}$ denotes an invertible normalization transformation. 
The suffix representations of an entire trajectory form $\mathbf{Y}_0=(\mathbf{y}_1,\ldots,\mathbf{y}_L)\in\mathbb{R}^{L\times d_y}$.
We employ conditional flow matching~\citep{lipman2023flowmatching} to learn a transport from Gaussian noise to the distribution of suffix latents. Given $\boldsymbol{\varepsilon}\sim\mathcal{N}(\mathbf{0},\mathbf{I})$, the interpolation at flow time $\eta\in[0,1]$ is defined as
\begin{equation}
\mathbf{Y}_\eta=(1-\eta)\mathbf{Y}_0+\eta\boldsymbol{\varepsilon},
\qquad
\mathbf{V}^{\star}=\boldsymbol{\varepsilon}-\mathbf{Y}_0.
\label{eq:path}
\end{equation}
A bidirectional Transformer $\mathcal{D}_\omega$ predicts the clean suffix latents $\hat{\mathbf{Y}}_0=\mathcal{D}_\omega(\mathbf{Y}_\eta,\eta,\mathbf{C})$, from which the velocity is estimated as $\hat{\mathbf{V}}=(\mathbf{Y}_\eta-\hat{\mathbf{Y}}_0)/\bar{\eta}$, where $\bar{\eta}=\max(\eta,10^{-6})$.
Following~\citet{li2026back}, we train the realizer by minimizing the velocity matching objective:
\begin{equation}
\mathcal{L}_{\mathrm{flow}}
=
\mathbb{E}_{\mathbf{Y}_0,\eta,\boldsymbol{\varepsilon}}
\left[
\left\|
\frac{\mathbf{Y}_\eta-\mathcal{D}_\omega(\mathbf{Y}_\eta,\eta,\mathbf{C})}{\bar{\eta}}
-\mathbf{V}^{\star}
\right\|_F^2
\right].
\label{eq:flow_loss}
\end{equation}
By jointly generating suffix latents conditioned on the complete plan, the realizer captures dependencies among fine-grained POI choices while preserving the planned coarse-grained activity structure.

\subsection{SID Decoding and POI Grounding}
\label{sec:grounding}

\paragraph{Coarse-to-fine generation.} SeMoFlow completes the generation in two stages:
(i) SID decoding selects a full Semantic ID consistent with
the planned prefix, and (ii) POI grounding maps the selected
SID to a concrete POI.
SeMoFlow first autoregressively generates a plan of visit
times and SID prefixes. Conditioned on this plan, the
realizer jointly generates suffix latents
$\mathbf{Y}^*=(\mathbf{y}_1^*,\ldots,\mathbf{y}_L^*)$
by transporting Gaussian noise through the learned reverse
flow.

\paragraph{Prefix-constrained SID decoding.}
For each visit, we consider only full SIDs in the target-city
catalog that share the planned prefix $\boldsymbol{\pi}_i$.
Each candidate is scored based on its latent similarity,
learned compatibility with the planned trajectory, and
frequency in the target-city training data
(see Appendix Eq.~(\ref{eq:decode_score})).
We then sample a full SID according to
\begin{equation}
\Pr(\hat{\mathbf z}_i=\mathbf z)
\propto
\exp\left(\frac{s_i(\mathbf z)}{T}\right),
\quad
\mathbf z\in\mathcal S_c(\boldsymbol{\pi}_i),
\label{eq:decode}
\end{equation}
where $\mathcal S_c(\boldsymbol{\pi}_i)$ denotes the
candidate SID set, $s_i(\mathbf z)$ is the candidate score,
and $T$ is the sampling temperature. The sampled SID is
then mapped to a POI, with collisions resolved when
multiple POIs share the same code. The detailed generation and decoding procedures
are provided in Appendix~\ref{app:generation}.

%% file: 05-Evaluation.tex
\section{Evaluation}\label{sec:evaluation}
In this section, we aim to address three questions: \textbf{RQ1:} How does \textit{SeMoFlow} compare with existing trajectory generation baselines? \textbf{RQ2:} How effective is \textit{SeMoFlow} in cross-city mobility generation and transfer? \textbf{RQ3:} How does each component contribute to the overall performance?

\subsection{Evaluation Setup}
\label{sec:evaluation_setup}

\noindent\textbf{Datasets:}
We evaluate SeMoFlow on two multi-city human
mobility datasets, Veraset and Foursquare. Each dataset contains anonymized visits with place categories, coordinates, and timestamps. 
For all experiments, we use device-disjoint
train/validation/test splits with an 80/10/10 ratio.
More details are presented in 
Table~\ref{tab:dataset_stats} (Appendix Sec.~\ref{app:dataset}).

\noindent\textbf{Implementation:}
We implement our \textit{SeMoFlow} and other baselines with PyTorch 2.12 and Python 3.12. 
All methods are trained with NVIDIA L40s GPUs (each with 46GB of memory) under a consistent hardware environment. 
The detailed settings of the semantic tokenizer, prefix planner, and flow-based denoising networks are summarized in Appendix Sec.~\ref{sec:Hyperparameters}.

\noindent\textbf{Metrics:}
To evaluate the quality and fidelity of the generated data, we first utilize our framework and the baselines to sample $30{,}000$ synthetic trajectories and then report six Jensen-Shannon divergences (JSD) between the synthetic datasets and the real dataset distributions:
(i) POI frequency ($JSD_{\mathrm{POI}}$), 
(ii) Geohash-5 spatial density ($JSD_{\mathrm{geo}}$), (iii) trajectory travel distance ($JSD_{\mathrm{travel}}$), 
(iv) radius of gyration ($JSD_{\mathrm{gyr}}$),
(v) daily transition-graph motifs ($JSD_{\mathrm{motif}}$)
~\citep{schneider2013unravelling}, 
and (vi) return-gap recurrence ($JSD_{\mathrm{recur}}$). 
Formal definitions of these metrics are given in
Appendix~\ref{app:metric_definitions}.

\noindent\textbf{Baselines:} 
To evaluate the effectiveness of our design, 
we compare it against the following trajectory generation baselines: 
(i) \textbf{Mechanistic and deep learning-based models}, including \textbf{TimeGeo}~\citep{jiang2016} and  \textbf{MoveSim}~\citep{feng2020movesim}, 
% (ii) \textbf{Semantic-agnostic diffusion-based methods}, i.e., \textbf{DiffTraj}~\citep{zhu2023difftraj}, 
and (ii) Semantic-aware POI sequence generative models (e.g., diffusion-based), including \textbf{MIRAGE}~\citep{deng2025mirage}, 
\textbf{Traveller}~\citep{luo2026traveller}, 
and \textbf{MobiDiff}~\citep{xu2026mobidiff}.
Details are elaborated in Appendix~\ref{app:Baselines}.

\subsection{Experimental Results}
\label{sec:results}

\subsubsection{Main Performance Comparison (RQ1)}
\label{sec:rq1}

\begin{table*}[t]
\centering
\caption{Comparison of trajectory generation performance across four cities.
Bold and underline indicate the best and second-best values within each city block, respectively.}
\label{tab:table1_baseline}

\scriptsize
\setlength{\tabcolsep}{2.2pt}
\renewcommand{\arraystretch}{0.9}

\begin{tabular*}{0.9\textwidth}{@{\extracolsep{\fill}}llcccccc@{}}
\toprule
\textbf{City} & \textbf{Model} &
$JSD_{\mathrm{POI}}\downarrow$ &
$JSD_{\mathrm{geo}}\downarrow$ &
$JSD_{\mathrm{travel}}\downarrow$ &
$JSD_{\mathrm{gyr}}\downarrow$ &
$JSD_{\mathrm{motif}}\downarrow$ &
$JSD_{\mathrm{recur}}\downarrow$ \\
\midrule

% ============================================================
% Chicago
% ============================================================
\multirow[c]{7}{*}{\textbf{Chicago}}

& MoveSim
& 0.1115 & 0.0087 & 0.0479 & 0.1346 & 0.4227 & 0.3399 \\

& TimeGeo
& 0.3286 & 0.0220 & 0.3240 & 0.2905 & 0.0810 & 0.0640 \\

& MIRAGE
& 0.1136 & 0.0091 & 0.0711 & 0.0709 & 0.0071 & 0.0257 \\

& MobiDiff
& 0.6187 & 0.4815 & \underline{0.0167} & 0.0265 & 0.0187 & 0.0275 \\

& Traveller
& 0.4096 & 0.1329 & 0.1909 & 0.2339 & 0.0037 & 0.0073 \\

\cmidrule(lr){2-8}

& \textit{SeMoFlow} (Per-City)
& \textbf{0.0943} & \textbf{0.0052} & 0.0172
& \textbf{0.0018} & \underline{0.0025} & \underline{0.0020} \\

& \textit{SeMoFlow} (Joint)
& \underline{0.1113}
& \underline{0.0060}
& \textbf{0.0158}
& \underline{0.0033}
& \textbf{0.0007}
& \textbf{0.0017} \\

\midrule

% ============================================================
% Boston
% ============================================================
\multirow[c]{7}{*}{\textbf{Boston}}

& MoveSim
& \underline{0.0971} & 0.0189 & 0.2334 & 0.2562 & 0.4466 & 0.2692 \\

& TimeGeo
& 0.3060 & 0.0326 & 0.3431 & 0.3218 & 0.0836 & 0.0674 \\

& MIRAGE
& 0.1424 & 0.0211 & 0.0609 & 0.0854 & 0.0495 & 0.0702 \\

& MobiDiff
& 0.5662 & 0.4201 & 0.0168 & 0.1428 & 0.0151 & 0.0306 \\

& Traveller
& 0.4054 & 0.1791 & 0.2032 & 0.3014 & 0.0036 & 0.0072 \\

\cmidrule(lr){2-8}

& \textit{SeMoFlow} (Per-City)
& \textbf{0.0796}
& \underline{0.0078}
& \textbf{0.0121}
& \underline{0.0046}
& \underline{0.0010}
& \underline{0.0023} \\

& \textit{SeMoFlow} (Joint)
& 0.0975
& \textbf{0.0077}
& \underline{0.0162}
& \textbf{0.0043}
& \textbf{0.0009}
& \textbf{0.0017} \\

\midrule

% ============================================================
% New York City
% ============================================================
\multirow[c]{7}{*}{\textbf{New York City}}

& MoveSim
& \textbf{0.1419} & 0.0160 & 0.0678 & 0.1055 & 0.1630 & 0.3881 \\

& TimeGeo
& 0.3584 & 0.0290 & 0.3668 & 0.3310 & 0.1279 & 0.1112 \\

& MIRAGE
& 0.1968 & 0.0147 & 0.1322 & 0.1718 & 0.0079 & 0.0489 \\

& MobiDiff
& 0.5881 & 0.3355 & \textbf{0.0202} & 0.0414 & 0.0193 & 0.0706 \\

& Traveller
& 0.4642 & 0.1823 & 0.0390 & 0.3817 & 0.0049 & 0.0280 \\

\cmidrule(lr){2-8}

& \textit{SeMoFlow} (Per-City)
& \underline{0.1443}
& \underline{0.0085}
& 0.0379
& \underline{0.0173}
& \underline{0.0027}
& \underline{0.0036} \\

& \textit{SeMoFlow} (Joint)
& 0.1463
& \textbf{0.0084}
& \underline{0.0206}
& \textbf{0.0044}
& \textbf{0.0008}
& \textbf{0.0030} \\

\midrule

% ============================================================
% Pittsburgh
% ============================================================
\multirow[c]{7}{*}{\textbf{Pittsburgh}}

& MoveSim
& 0.0907 & 0.0201 & 0.2758 & 0.2971 & 0.4650 & 0.2179 \\

& TimeGeo
& 0.2913 & 0.0386 & 0.3429 & 0.3409 & 0.0842 & 0.0570 \\

& MIRAGE
& 0.1242 & 0.0194 & 0.0554 & 0.0915 & 0.0490 & 0.0718 \\

& MobiDiff
& 0.5715 & 0.4329 & 0.0628 & 0.0530 & 0.0440 & 0.0287 \\

& Traveller
& 0.3760 & 0.1640 & 0.1872 & 0.2572 & 0.0060 & 0.0063 \\

\cmidrule(lr){2-8}

& \textit{SeMoFlow} (Per-City)
& \textbf{0.0648}
& \underline{0.0080}
& \underline{0.0396}
& \underline{0.0394}
& \underline{0.0031}
& \underline{0.0017} \\

& \textit{SeMoFlow} (Joint)
& \underline{0.0715}
& \textbf{0.0072}
& \textbf{0.0174}
& \textbf{0.0114}
& \textbf{0.0012}
& \textbf{0.0011} \\

\bottomrule
\end{tabular*}
\end{table*}

We first evaluate SeMoFlow by training and testing each model separately on four cities. As shown in Table~\ref{tab:table1_baseline}, SeMoFlow achieves competitive generation fidelity across all cities, particularly in preserving spatial distributions and long-term mobility patterns. Notably, it achieves the lowest recurrence divergence in all four cities, while maintaining strong performance in POI frequency and geographic fidelity. Additional visualizations of mobility flows, spatial and
activity distributions, and motif patterns are provided
in Appendix Secs.~\ref{sec:visualization}
and~\ref{sec:Recurrence Patterns}.

We further train a single SeMoFlow model jointly on all four cities and evaluate its performance separately in each city. Unlike existing baselines that require city-specific POI representations and separate training, SeMoFlow enables joint generation through a unified Semantic ID space shared across heterogeneous cities. 
The joint model maintains competitive fidelity in most cities and improves several mobility metrics in New York City and Pittsburgh, demonstrating the potential of shared semantic representations for multi-city mobility generation. 
Additional results on the Foursquare dataset are
reported in Appendix Sec.~\ref{sec:table_foursquare}.

\subsubsection{Multi-City Generation and Transfer Performance (RQ2)}
\label{sec:rq2}

\begin{figure}[h]
    \centering
    \includegraphics[width=0.95\linewidth]{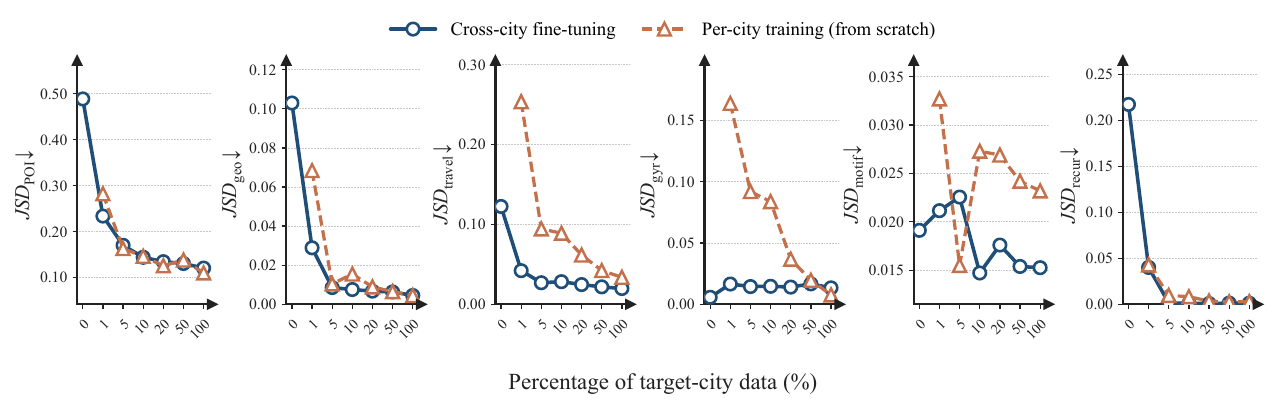}
    \vspace{-5mm}
    \caption{Cross-city adaptation to Dallas using different percentages of target-city data.}
    \label{fig:cross-city}
\end{figure}

To assess cross-city transferability, we evaluate SeMoFlow on Dallas, a city excluded from source training, and compare fine-tuning the joint model with training from scratch.
For cross-city transfer, SeMoFlow has access to the
target-city POI catalog and metadata, while held-out mobility sequences remain unseen.
As shown in Fig.~\ref{fig:cross-city}, cross-city transfer
substantially improves geographic fidelity, travel distance,
and radius of gyration when target-city data are limited. 
The performance gap gradually narrows as more Dallas data
become available. 
These results demonstrate the effectiveness of cross-city
knowledge transfer for data-efficient mobility generation.

\subsubsection{Ablation Studies (RQ3)}
\label{sec:rq3}

We conduct ablation studies on Chicago to examine the contribution of each component. Specifically, w/o Graph removes the POI transition graph embeddings from SID tokenization; w/o Planner models the full SID representation through latent flow matching without hierarchical prefix planning; and w/o Router removes the Explore-Retrieve mechanism from the autoregressive planner.

\begin{table}[h]
\centering
\caption{Ablation studies of \textit{SeMoFlow} on Chicago dataset. Bold indicates the best result.}
\label{tab:ablation}
\scriptsize
\setlength{\tabcolsep}{7pt}
\renewcommand{\arraystretch}{0.8}

\begin{tabular}{lcccccc}
\toprule

\textbf{Variant} &
$JSD_{\mathrm{POI}}\downarrow$ &
$JSD_{\mathrm{geo}}\downarrow$ &
$JSD_{\mathrm{travel}}\downarrow$ &
$JSD_{\mathrm{gyr}}\downarrow$ &
$JSD_{\mathrm{motif}}\downarrow$ &
$JSD_{\mathrm{recur}}\downarrow$ \\
\midrule

w/o Graph
& 0.0978 & \textbf{0.0049} & 0.2133
& 0.2798 & 0.0026 & 0.0056 \\

w/o Planner
& 0.4093 & 0.0409 & 0.0745
& 0.0304 & 0.0055 & 0.0052 \\

w/o Router
& \textbf{0.0943} & 0.0054 & 0.0184
& 0.0019 & 0.0048 & 0.0060 \\

\midrule
\textbf{\textit{SeMoFlow}}
& \textbf{0.0943} & 0.0052 & \textbf{0.0172}
& \textbf{0.0018} & \textbf{0.0025} & \textbf{0.0020} \\

\bottomrule
\end{tabular}
\end{table}

As shown in Table~\ref{tab:ablation}, removing graph embeddings substantially increases travel-distance and gyration divergences. Removing the planner leads to a marked deterioration in POI distribution and geographic fidelity, while removing the router increases recurrence divergence from 0.0020 to 0.0060. These results support the complementary roles of these sub-module designs.

\begin{figure}[t]
    \centering
    \includegraphics[width=0.9\linewidth]{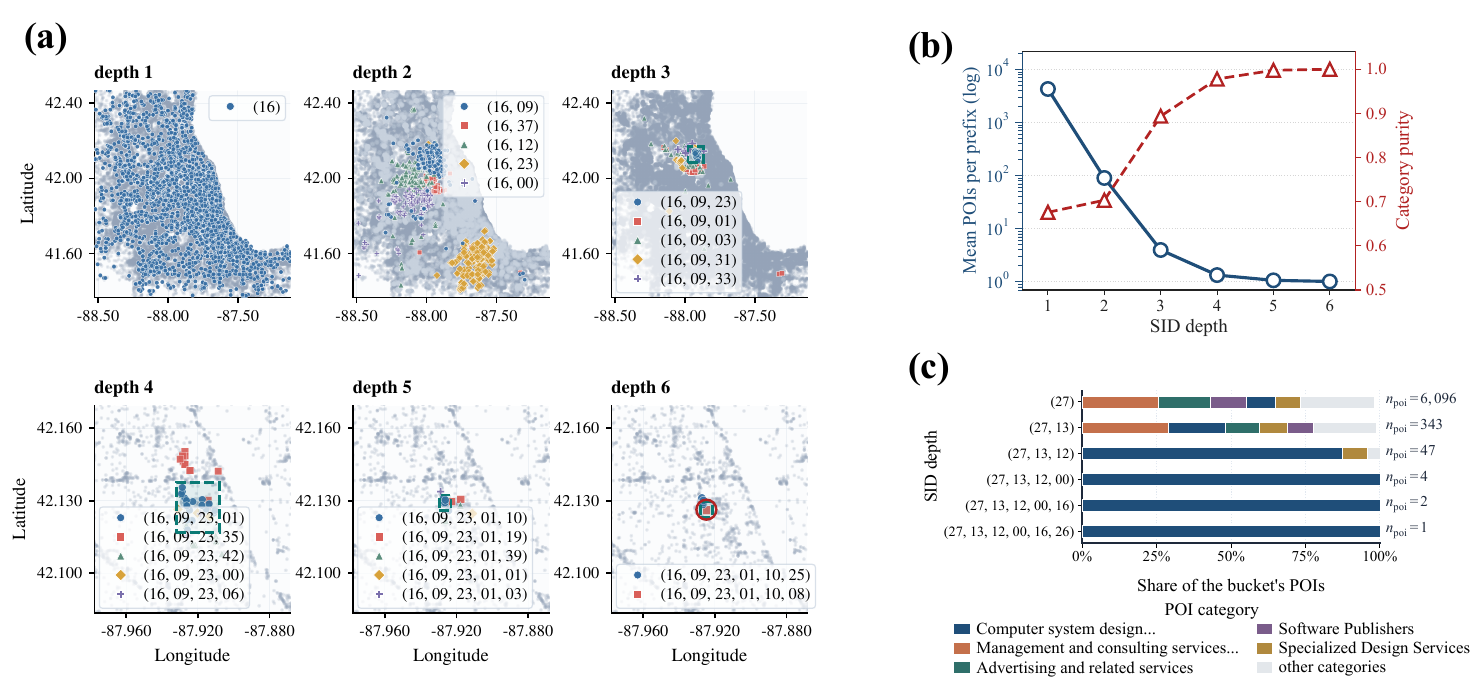}
    \caption{
    Hierarchical SIDs in Chicago. 
    (a) Coarse-to-fine spatial distributions across six SID levels. (b) Mean POIs per prefix and category purity at each depth. (c) An illustrative example in which the prefix captures a dominant POI category and subsequent
codes progressively distinguish individual POIs.}
    \label{fig:sid}
    \vspace{-5mm}
\end{figure}

\subsubsection{Interpretability of Semantic IDs}
Fig.~\ref{fig:sid} illustrates the hierarchical structure learned by the POI tokenizer. As shown in Fig.~\ref{fig:sid}(a), successive SID levels progressively refine the spatial distribution of POIs, from broad urban regions to individual locations. Fig.~\ref{fig:sid}(b) reports the mean number of POIs per prefix and category purity, defined as the POI-weighted proportion of the dominant category within each prefix group. The first three codes capture distinctive POI semantics, while subsequent codes progressively reduce the number of POIs sharing the same prefix. Fig.~\ref{fig:sid}(c) provides an example in which a three-level prefix corresponds to a specific activity category, with the remaining codes distinguishing individual POIs.

%% file: 06-Conclusion.tex
\section{Conclusion}\label{sec:conclusion} 

In this paper, we present SeMoFlow, a coarse-to-fine framework for generating human POI visitation trajectories across cities. SeMoFlow constructs hierarchical Semantic IDs that map city-specific POIs into a unified semantic representation space. 
Building on this representation, an autoregressive prefix
planner captures recurring visitation patterns, while a
conditional flow-matching realizer synthesizes fine-grained
suffix latents that are subsequently decoded and grounded
to POIs.
Experiments on two multi-city mobility datasets demonstrate SeMoFlow's effectiveness in trajectory generation and cross-city transfer.

%% file: 07-Appendix.tex
\section{Coarse-to-fine Trajectory Generation}
\label{app:generation}

Algorithm~\ref{alg:generation} summarizes the generation procedure of SeMoFlow, from autoregressive prefix planning and conditional flow sampling to prefix-constrained SID decoding and POI grounding.

\begin{algorithm}[h]
\caption{Coarse-to-fine POI Trajectory Generation with SeMoFlow}
\label{alg:generation}
\begin{algorithmic}[1]
\Require Target city $c$; planner $p_\theta$;
realizer $F_{\mathrm{cond}},\mathcal{D}_\omega$;
POI catalog $\mathcal{P}_c$; Euler steps $N_{\mathrm E}$.

\State Sample trajectory length $L\sim p_\theta(L\mid c)$;
$\bar t_0\gets 0$; $\mathcal M_1\gets\varnothing$.
\Comment{Coarse-grained trajectory planning}
\For{$i=1,\ldots,L$}
    \State Compute $\mathbf h_i$;
    sample $g_i\sim p_\theta(g_i\mid\mathbf h_i)$.
    \State $\bar t_i\gets
    \min(\Omega,\bar t_{i-1}+\bar\delta(g_i))$;
    $q_i\gets\mathrm{bin}(\bar t_i)$.
    \State Compute $\tilde{\mathbf h}_i$ and gate $r_i$;
    set $r_i\gets 0$ if $\mathcal M_i=\varnothing$.
    \State Sample SID prefix $\boldsymbol\pi_i$
    via Explore--Retrieve according to $r_i$;
    update $\mathcal M_{i+1}$.
\EndFor

\State $\mathbf C\gets
F_{\mathrm{cond}}(\boldsymbol\pi_{1:L},q_{1:L},c)$;
$\mathbf Y\sim\mathcal N(\mathbf 0,\mathbf I)$.
\Comment{Joint suffix realization}
\For{$n=N_{\mathrm E},\ldots,1$}
    \State $\eta\gets n/N_{\mathrm E}$;
    $\mathbf Y\gets\mathbf Y-
    \frac{1}{N_{\mathrm E}}
    \frac{\mathbf Y-\mathcal D_\omega
    (\mathbf Y,\eta,\mathbf C)}{\eta}$.
\EndFor

\State $\mathbf Y^*\gets\mathbf Y$;
sample each full SID $\hat{\mathbf z}_i$
under $\boldsymbol\pi_i$
using Eq.~(\ref{eq:decode}).
\Comment{Prefix-constrained SID decoding}
\State Map $\hat{\mathbf z}_{1:L}$
to POIs $\hat p_{1:L}$, resolving SID collisions.
\Comment{POI grounding}
\State \Return POI trajectory
$\hat\tau=\{(\bar t_i,\hat p_i)\}_{i=1}^{L}$.
\end{algorithmic}
\end{algorithm} 
Here, $\bar{\delta}(g_i)$ maps the sampled gap token $g_i$
to its representative duration in minutes, $\bar t_i$
denotes the accumulated visit time from the beginning of
the week, and $q_i=\mathrm{bin}(\bar t_i)$ denotes the
corresponding weekly time token. We discretize inter-visit
gaps into 16 bins over a one-week horizon, while the 42
time tokens encode seven weekdays and six within-day time
periods. The maximum trajectory horizon is
$\Omega=10080$ minutes.

\paragraph{SID decoding.}
For each planned prefix $\boldsymbol{\pi}_i$, the decoder
considers only the full SIDs in the target-city catalog
that extend this prefix. Each candidate
$\mathbf z\in\mathcal S_c(\boldsymbol{\pi}_i)$
is assigned a score:
\begin{equation}
s_i(\mathbf z)
=
\underbrace{a_i(\mathbf z)}_{\text{Compatibility}}
-
\underbrace{
\frac{1}{2d_y}
\left\|\mathbf y_i^*-\tilde{\mathbf y}_{\mathbf z}\right\|_2^2
}_{\text{Latent distance}}
+
\underbrace{
\lambda\log\bigl(1+n_c(\mathbf z)\bigr)
}_{\text{Catalog prior}}.
\label{eq:decode_score}
\end{equation}
where (i) \textit{Compatibility} $a_i(\mathbf z)$ measures
how well the candidate SID matches the current visit's
planned context and generated suffix latent;
(ii) \textit{Latent distance} encourages
consistency between the generated suffix latent and
the candidate's suffix representation, where
$\tilde{\mathbf y}_{\mathbf z}$ denotes the normalized
sum of its suffix codebook vectors;
and (iii) \textit{Catalog prior} favors SIDs frequently
visited in the target city, where $n_c(\mathbf z)$
denotes the occurrence count of the full SID in the
target-city training split.
The resulting scores
define the sampling distribution in Eq.~(\ref{eq:decode}). 
The learned compatibility head is trained with a
cross-entropy objective to identify the ground-truth
SID among valid candidates sharing the same prefix,
using the realizer's predicted suffix latents.

\paragraph{POI grounding.}
SID decoding selects a full code rather than a unique POI. We map each sampled SID to a concrete POI using the canonical first-candidate rule, which selects the first POI associated with that SID in the catalog when multiple POIs share the same code.

\section{Detailed Experimental Setup}
\label{app:setup}

\subsection{Dataset descriptions}
\label{app:dataset}
We evaluate SeMoFlow on two multi-city human mobility datasets, Veraset and Foursquare. Their statistics and experimental settings are summarized in Table~\ref{tab:dataset_stats}. 
\begin{itemize}[leftmargin=*]
    \item \textbf{Veraset.} We use large-scale mobile device mobility data from Veraset~\citep{veraset2022visits}, accessed through Dewey Data\footnote{\url{https://www.deweydata.io/data-partners/veraset}}. The dataset covers five U.S. metropolitan areas: Boston, Chicago, New York City, Pittsburgh, and Dallas. The first four cities are used for per-city and joint multi-city generation, while Dallas is held out from source training to evaluate cross-city transferability. 
    \item \textbf{Foursquare.} We further evaluate SeMoFlow on the Foursquare check-in dataset\footnote{\url{https://sites.google.com/site/yangdingqi/home/foursquare-dataset}}~\citep{yang2016participatory}, which records user visits to POIs with geographic coordinates, timestamps, and activity categories. We select three U.S. cities: Los Angeles, Washington, DC, and Atlanta, to examine the generation performance on an additional mobility data source.
\end{itemize}
 \paragraph{Data preprocessing.} For both datasets, we organize individual visits into weekly trajectories and retain sequences containing 5 to 128 visits. Each trajectory consists of chronologically ordered POI visits with their corresponding timestamps and locations. Dataset statistics, including the number of POIs and weekly trajectory segments, are reported in Table~\ref{tab:dataset_stats}. 

\paragraph{Data split.}
We adopt device-disjoint rather than chronological data
splits for all experiments. Devices are assigned to
training, validation, and test sets with an 80/10/10 ratio
based on a deterministic hash of their anonymized device
identifiers (CAIDs), ensuring that all trajectories from
the same device remain within a single split.

\begin{table*}[h]
\centering
\caption{Two multi-city human mobility datasets, i.e., Veraset and Foursquare check-in datasets.}
\label{tab:dataset_stats}

\scriptsize
\setlength{\tabcolsep}{2.2pt}
\renewcommand{\arraystretch}{0.9}

\begin{tabular*}{0.95\textwidth}{@{\extracolsep{\fill}}llcccc@{}}
\toprule
\textbf{Dataset} &
\textbf{City} &
\textbf{\# POIs} &
\textbf{\# Weekly Segments} &
\textbf{Length Range ($L$)} &
\textbf{Experimental Settings} \\
\midrule

\multirow[c]{5}{*}{\textbf{Veraset}}
& Boston
& 122,731
& 390,148
& $5 \le L \le 128$
& Per-City / Multi-City training \\

& Chicago
& 206,852
& 679,417
& $5 \le L \le 128$
& Per-City / Multi-City training \\

& New York City
& 394,469
& 1,609,108
& $5 \le L \le 128$
& Per-City / Multi-City training \\

& Pittsburgh
& 68,869
& 219,842
& $5 \le L \le 128$
& Per-City / Multi-City training \\

& Dallas
& 180,268
& 939,831
& $5 \le L \le 128$
& Only for Cross-City Transfer\\

\midrule

\multirow[c]{4}{*}{\textbf{Foursquare}}
& Los Angeles
& 25,889
& 10,197
& $5 \le L \le 128$
& Per-City / Multi-City training \\

& Washington, DC
& 19,416
& 9,196
& $5 \le L \le 128$
& Per-City / Multi-City training \\

& Atlanta
& 13,963
& 6,472
& $5 \le L \le 128$
& Per-City / Multi-City training \\

\bottomrule
\end{tabular*}
\end{table*}

\subsection{Training and Evaluation Settings}
\label{sec-app:training-setting}

We consider three training and evaluation settings.

\textbf{Per-city training.}
A separate SeMoFlow model is trained on the training data
of each city and evaluated on the held-out trajectories
from the same city.

\textbf{Multi-city training with per-city testing.}
A single SeMoFlow model is jointly trained on multiple
cities using the shared Semantic ID space. The jointly
trained model is then evaluated separately on the held-out
trajectories of each city.

\textbf{Cross-city transfer.}
For the Veraset experiments, Boston, Chicago, New York City,
and Pittsburgh are used as source cities, while Dallas is
held out from source training. The jointly trained source
model is adapted to Dallas using different fractions of
target-city training data and compared with models trained
from scratch using the same data budget. SeMoFlow has
access to the target-city POI catalog, including POI
metadata and geographic information, while held-out
mobility sequences remain unseen during training and
adaptation. At the 0\% setting, no target-city mobility
observations or visitation priors are used.

\subsection{Implementation Details} \label{sec:Hyperparameters} Table~\ref{tab:architecture} summarizes the architecture hyperparameters of SeMoFlow, including the hierarchical SID tokenizer, autoregressive prefix planner, and latent flow matching realizer. Table~\ref{tab:training_hyperparameters} provides the training configurations for both per-city and joint multi-city settings, including learning rates, batch sizes, training epochs, and other key hyperparameters.

% Dallas adaptation implementation details

\begin{table*}[t]
\centering
\caption{Architecture hyperparameters of SeMoFlow.}
\label{tab:architecture}

\scriptsize
\setlength{\tabcolsep}{4pt}
\renewcommand{\arraystretch}{1.15}

\begin{tabular*}{0.95\textwidth}{
@{\extracolsep{\fill}}ll|ll|ll@{}
}
\hline

\multicolumn{2}{c|}{\textbf{SID Tokenizer}} &
\multicolumn{2}{c|}{\textbf{Prefix Planner}} &
\multicolumn{2}{c}{\textbf{Latent-Flow Realizer}} \\
\hline

\textbf{Parameter} & \textbf{Value} &
\textbf{Parameter} & \textbf{Value} &
\textbf{Parameter} & \textbf{Value} \\
\hline

SID levels & 6 &
Depth & 6 &
Transformer width & 256 \\

Codebook size & 32 (Boston / Pittsburgh) &
Attention heads & 8 &
Latent width & 256 \\

Codebook size & 48 (Chicago / NYC) &
Gap bins & 16 &
Attention heads & 8 \\

Prefix levels & 1--3 &
Time bins & 42 &
Flow trunk & 6 layers \\

Suffix levels & 4--6 &
Max. occurrences & 128 &
Max. occurrences & 128 \\

& &
Dropout & 0.1 &
Dropout & 0.1 \\

\hline
\end{tabular*}
\end{table*}

\begin{table*}[t]
\centering
\caption{Training hyperparameters of SeMoFlow.}
\label{tab:training_hyperparameters}

\scriptsize
\setlength{\tabcolsep}{4pt}
\renewcommand{\arraystretch}{1.15}
\setlength{\extrarowheight}{2pt}

% Vertically center the X column.
\renewcommand{\tabularxcolumn}[1]{m{#1}}

\begin{tabularx}{0.95\textwidth}{
@{}
m{0.09\textwidth}|
m{0.14\textwidth}
m{0.31\textwidth}|
m{0.14\textwidth}
X
@{}
}
\hline

\textbf{Module} &
\multicolumn{2}{c|}{\textbf{Per-City Training}} &
\multicolumn{2}{c}{\textbf{Joint Training}} \\
\hline

& \textbf{Parameter} & \textbf{Value} &
\textbf{Parameter} & \textbf{Value} \\
\hline

\multirow[c]{5}{*}{\textbf{Planner}}
& Learning rate
& \makecell[l]{
$1.4\times10^{-3}$ (Boston/NYC/Pittsburgh)\\[2pt]
$2\times10^{-3}$ (Chicago)}
& Learning rate
& $1.4\times10^{-3}$ \\
\cline{2-5}

& Batch size
& \makecell[l]{
1024 (Boston); 4096 (Chicago)\\[2pt]
2048 (NYC/Pittsburgh)}
& Batch size
& 2304 \\
\cline{2-5}

& Epochs
& \makecell[l]{
30 (Boston/Pittsburgh)\\[2pt]
50 (Chicago); 60 (NYC)}
& Epochs
& 100 \\
\cline{2-5}

& Weight decay
& $10^{-4}$
& Weight decay
& $10^{-4}$ \\
\cline{2-5}

& Routing temperature
& 1.7
& Routing temperature
& 2.0 \\

\hline

\multirow[c]{5}{*}{\textbf{Realizer}}
& Learning rate
& \makecell[l]{
$2\times10^{-4}$ (Boston/Chicago/Pittsburgh)\\[2pt]
$5\times10^{-5}$ (NYC fine-tuning)}
& Learning rate
& $2\times10^{-4}$ \\
\cline{2-5}

& Batch size
& \makecell[l]{
512 (Boston/Chicago)\\[2pt]
1024 (NYC/Pittsburgh)}
& Batch size
& 768 \\
\cline{2-5}

& Epochs
& \makecell[l]{
100 (Boston/Chicago/Pittsburgh)\\[2pt]
30 (NYC fine-tuning)}
& Epochs
& 100 \\
\cline{2-5}

& Weight decay
& 0.01
& Weight decay
& 0.01 \\
\cline{2-5}

& Catalog temperature
& 0.8
& Catalog temperature
& \makecell[l]{
0.45 (Boston)\\[2pt]
0.6 (other cities)} \\

\hline
\end{tabularx}
\end{table*}

\subsection{Metric Computation}
\label{app:metric_definitions}

We evaluate trajectory fidelity by comparing the empirical distributions of six mobility statistics between real ($\mathcal{R}$) and generated ($\mathcal{G}$) trajectories. All distributional metrics use the Jensen--Shannon divergence (JSD):
\begin{equation}
D_{\mathrm{JS}}(P,Q)
=\frac{1}{2}D_{\mathrm{KL}}(P\|M)
+\frac{1}{2}D_{\mathrm{KL}}(Q\|M),
\quad M=\frac{P+Q}{2}.
\label{eq:jsd}
\end{equation}
Lower divergence indicates closer agreement with real mobility.

\begin{itemize}[leftmargin=*]

\item \textbf{POI frequency ($JSD_{\mathrm{POI}}$).}
We count visits to each POI in the city catalog and compute the JSD between the normalized real and generated visit-frequency distributions.

\item  \textbf{Spatial density ($JSD_{\mathrm{geo}}$).}
Each POI visit is mapped to its Geohash-5 cell. We compare the normalized visit-frequency distributions over the union of spatial cells.

\item \textbf{Travel distance ($JSD_{\mathrm{travel}}$).}
For each trajectory, we compute the total Haversine distance between consecutive visits with valid coordinates. Visits with missing coordinates are skipped, connecting the nearest available neighbors. The resulting trajectory distances are discretized into 50 equal-width bins over the combined real and generated range, and their distributions are compared using JSD.

\item \textbf{Radius of gyration ($JSD_{\mathrm{gyr}}$).}
For a trajectory with $L$ valid visit coordinates $\{\mathbf{x}_i\}_{i=1}^{L}$, its radius of gyration is
\begin{equation}
r_g=\sqrt{\frac{1}{L}\sum_{i=1}^{L}
d_H(\mathbf{x}_i,\bar{\mathbf{x}})^2},
\label{eq:radius_gyration}
\end{equation}
where $d_H$ denotes Haversine distance and $\bar{\mathbf{x}}$ is the coordinate-wise mean of the visited locations. Trajectories with fewer than two valid coordinates are assigned zero radius. We compare the real and generated radius distributions using 50 equal-width bins over their combined range.

\item \textbf{Daily motifs ($JSD_{\mathrm{motif}}$).}
Following~\citet{schneider2013unravelling}, we represent each daily trajectory as a directed, unweighted graph of visited POIs. Consecutive duplicate visits are collapsed, and locations are labeled by their first-visit order to construct a canonical motif signature. Motifs occurring in more than $0.5\%$ of real trajectory-days are retained, while all remaining motifs are grouped into an \emph{other} category. We compute the JSD between the resulting daily motif distributions.

\item \textbf{Recurrence ($JSD_{\mathrm{recur}}$).}
For each repeated POI visit at position $i$, we define its return gap as $\Delta_i=i-j$, where $j<i$ is the most recent preceding visit to the same POI. Thus, consecutive visits to the same POI yield a gap of one. We compare the real and generated distributions of return gaps using JSD.

\end{itemize}

\subsection{Baselines}
\label{app:Baselines}

We compare SeMoFlow with five representative baselines,
covering two categories: (i) mechanism-driven and
adversarial mobility models, and (ii) diffusion-based
generative models.

\begin{itemize}[leftmargin=*]

\item \textbf{TimeGeo}~\citep{jiang2016} is a
mechanism-driven mobility model that captures temporal
routines and spatial visitation patterns. It models
individual activities through home-based tours,
location selection, and preferential returns.

\item \textbf{MoveSim}~\citep{feng2020movesim} employs
adversarial sequence generation to synthesize human
mobility trajectories. It incorporates mobility priors,
including historical transitions and spatial
relationships, to guide trajectory generation.

\item \textbf{MIRAGE}~\citep{deng2025mirage} combines
a neural temporal point process with exploration and
preferential-return mechanisms. It explicitly models
the choice between exploring new locations and
revisiting previously visited ones.

\item \textbf{Traveller}~\citep{luo2026traveller}
adopts a hierarchical generative framework combining
autoregressive temporal planning with diffusion-based
spatial generation. It extracts individual travel
patterns to condition the generation of mobility
trajectories.

\item \textbf{MobiDiff}~\citep{xu2026mobidiff} is a
discrete diffusion model for semantic human mobility
generation. It jointly models spatial, temporal,
and activity semantics through structured denoising
of discrete mobility sequences.

\end{itemize}

\section{Additional Experimental Results}

\subsection{Generated Trajectory Data Visualization}
\label{sec:visualization}

\begin{figure}[h]
    \centering
    \includegraphics[width=1\linewidth]{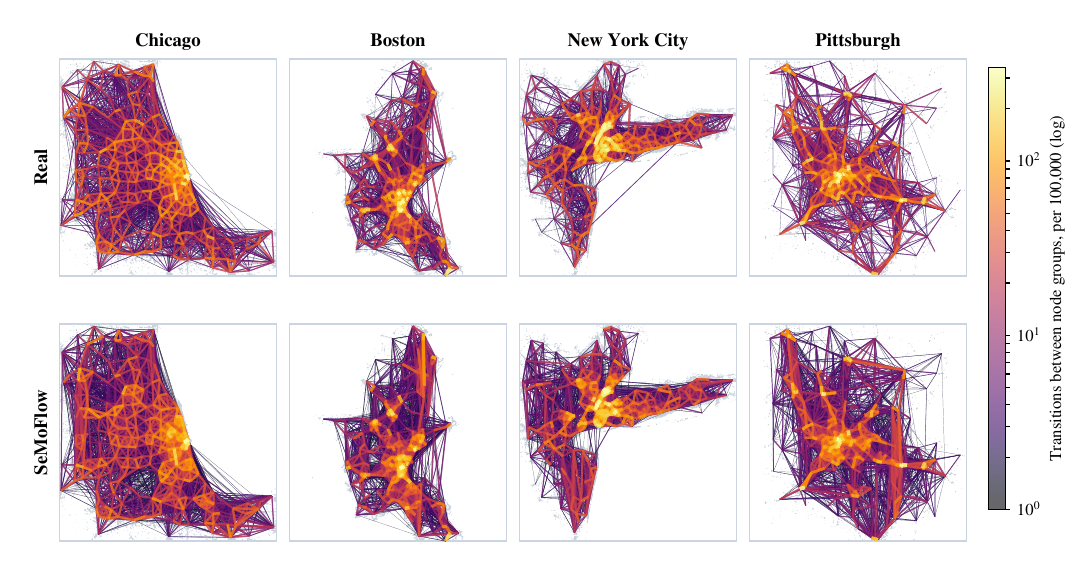}
    \caption{Mobility flow networks of real and SeMoFlow-generated trajectories across four cities.}
    \label{fig:flow_graph}
\end{figure}

\paragraph{Mobility flow networks of real and synthetic data:} As shown in Fig.~\ref{fig:flow_graph}, SeMoFlow reproduces the major mobility hubs, frequently traveled corridors, and overall network coverage across all four cities. The generated networks closely resemble the real flow structures, with remaining discrepancies primarily observed in less frequent transitions between peripheral locations. These results indicate that SeMoFlow captures not only individual POI visitation patterns but also the spatial connectivity underlying human mobility.

\begin{figure}[h]
    \centering
    \includegraphics[width=1\linewidth]{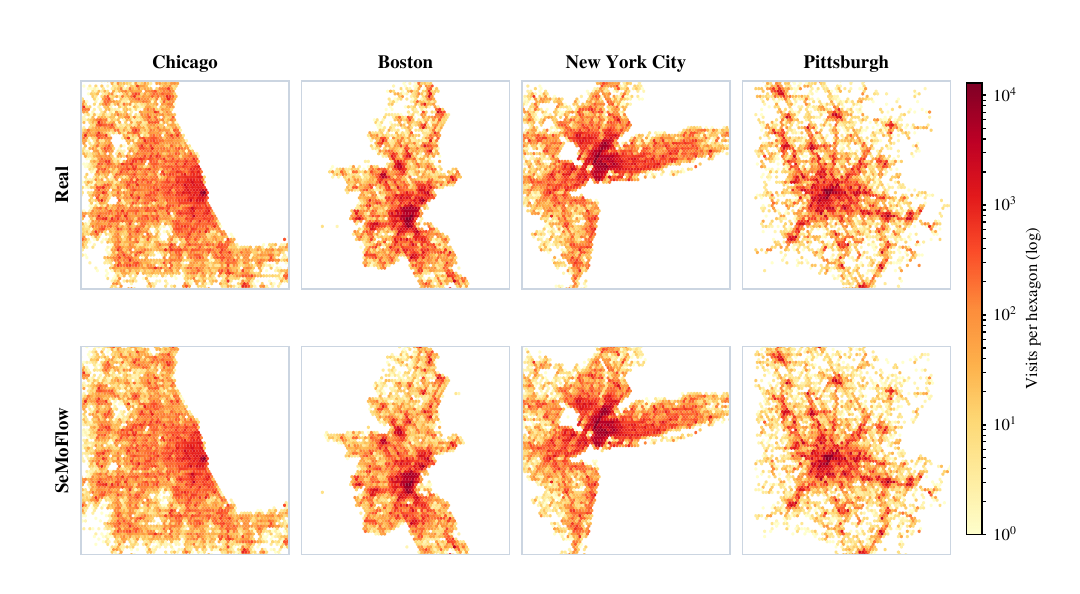}
    \caption{Spatial distributions of real and SeMoFlow-generated POI visits across four cities. Visit frequencies are aggregated into hexagonal spatial bins and visualized using a shared logarithmic color scale.}
    \label{fig:activity_density}
\end{figure}

\paragraph{Spatial distributions of real and SeMoFlow-generated POI visits across four cities:}
Fig.~\ref{fig:activity_density} shows that SeMoFlow closely reproduces the spatial distribution of human activities, including major activity centers, high-density corridors, and the overall geographic extent of visits. This spatial agreement suggests that the hierarchical prefix--suffix representation preserves fine-grained location preferences while maintaining the broader spatial structure of urban mobility.

\begin{figure}[h]
    \centering
    \includegraphics[width=1\linewidth]{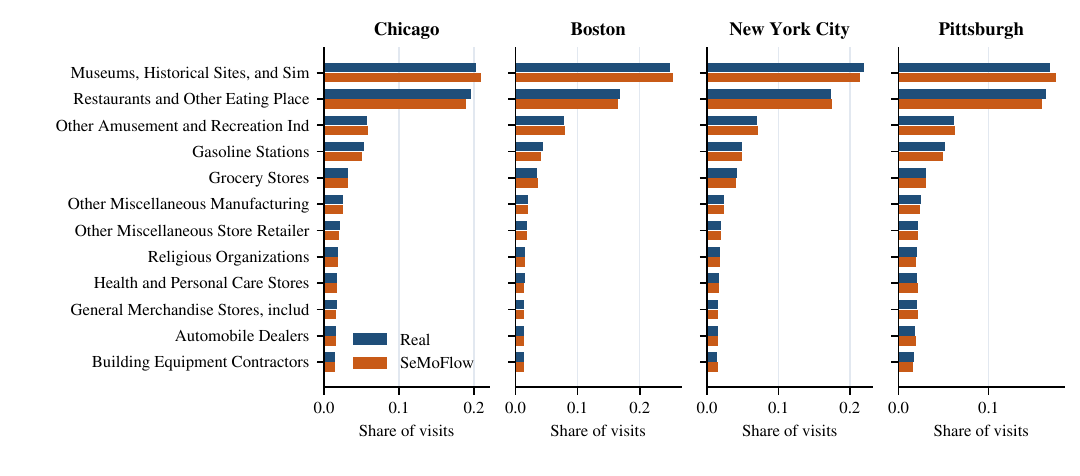}
    \caption{POI category distributions of real and SeMoFlow-generated trajectories across four cities. Bars compare the share of visits to the twelve most frequently visited POI categories in each city.}
    \label{fig:activity_mix}
\end{figure}

\paragraph{POI category distributions of real and SeMoFlow-generated trajectories across four cities:}
As shown in Fig.~\ref{fig:activity_mix}, SeMoFlow closely matches the real POI category distributions across all four cities. The generated visit proportions are consistent with the observed activity composition, from dominant categories such as restaurants and recreational venues to less frequent categories accounting for fewer than 2\% of visits. This agreement provides further evidence that the learned Semantic IDs retain functional POI information during trajectory generation.

\subsection{Performance comparison in Foursquare Dataset}
\label{sec:table_foursquare}

\begin{table*}[h]
\centering
\caption{Trajectory generation comparison across three Foursquare cities.
Bold and underline indicate the best and second-best values within each city block,
respectively.}
\label{tab:table2_foursquare}

\scriptsize
\setlength{\tabcolsep}{2.2pt}
\renewcommand{\arraystretch}{0.9}

\begin{tabular*}{0.95\textwidth}{@{\extracolsep{\fill}}llcccccc@{}}
\toprule
\textbf{City} & \textbf{Model} &
$JSD_{\mathrm{POI}}\downarrow$ &
$JSD_{\mathrm{geo}}\downarrow$ &
$JSD_{\mathrm{travel}}\downarrow$ &
$JSD_{\mathrm{gyr}}\downarrow$ &
$JSD_{\mathrm{motif}}\downarrow$ &
$JSD_{\mathrm{recur}}\downarrow$ \\
\midrule

% ============================================================
% Los Angeles
% ============================================================
\multirow[c]{7}{*}{\textbf{Los Angeles}}

& MoveSim
& 0.5282
& 0.0467
& 0.4850
& 0.4333
& 0.5963
& 0.4274 \\

& TimeGeo
& 0.5558
& \underline{0.0424}
& \underline{0.0597}
& \underline{0.0683}
& 0.0311
& 0.0720 \\

& MIRAGE
& \textbf{0.4729}
& 0.0462
& 0.1367
& 0.1841
& \underline{0.0020}
& 0.0182 \\

& MobiDiff
& 0.6144
& 0.3996
& 0.0945
& 0.1795
& 0.0782
& 0.2363 \\

& Traveller
& 0.5596
& 0.0635
& 0.1496
& 0.1538
& 0.0037
& \textbf{0.0064} \\

\cmidrule(lr){2-8}

& \textit{SeMoFlow} (Per-City)
& 0.4883
& 0.0564
& 0.3968
& 0.1474
& 0.0629
& 0.1072 \\

& \textit{SeMoFlow} (Joint)
& \underline{0.4828}
& \textbf{0.0406}
& \textbf{0.0262}
& \textbf{0.0382}
& \textbf{0.0005}
& \underline{0.0170} \\

\midrule

% ============================================================
% Washington, DC
% ============================================================
\multirow[c]{7}{*}{\textbf{Washington, DC}}

& MoveSim
& 0.4924
& \textbf{0.0476}
& 0.4359
& 0.3706
& 0.5758
& 0.1521 \\

& TimeGeo
& 0.5361
& \underline{0.0496}
& \underline{0.0562}
& \underline{0.0451}
& 0.0272
& 0.0531 \\

& MIRAGE
& \underline{0.4573}
& 0.0670
& 0.0613
& 0.1006
& \underline{0.0006}
& \underline{0.0105} \\

& MobiDiff
& 0.6080
& 0.2340
& 0.0705
& 0.2310
& 0.0432
& 0.0652 \\

& Traveller
& 0.5581
& 0.0786
& 0.1293
& 0.1844
& 0.0019
& \textbf{0.0081} \\

\cmidrule(lr){2-8}

& \textit{SeMoFlow} (Per-City)
& 0.4937
& 0.1040
& 0.3550
& 0.1687
& 0.0749
& 0.1188 \\

& \textit{SeMoFlow} (Joint)
& \textbf{0.4517}
& 0.0530
& \textbf{0.0267}
& \textbf{0.0330}
& \textbf{0.0004}
& 0.0235 \\

\midrule

% ============================================================
% Atlanta
% ============================================================
\multirow[c]{7}{*}{\textbf{Atlanta}}

& MoveSim
& 0.4996
& \textbf{0.0723}
& 0.4717
& 0.3195
& 0.5739
& 0.1829 \\

& TimeGeo
& 0.5462
& \underline{0.0775}
& \underline{0.0602}
& \underline{0.0364}
& 0.0298
& 0.0708 \\

& MIRAGE
& \textbf{0.4674}
& 0.0897
& 0.0715
& 0.0961
& 0.0071
& 0.0186 \\

& MobiDiff
& 0.6278
& 0.4258
& 0.0965
& 0.1746
& 0.0698
& 0.2198 \\

& Traveller
& 0.5645
& 0.0966
& 0.1182
& 0.1377
& \textbf{0.0020}
& \textbf{0.0095} \\

\cmidrule(lr){2-8}

& \textit{SeMoFlow} (Per-City)
& 0.4799
& 0.1001
& 0.3888
& 0.1763
& 0.0539
& 0.0862 \\

& \textit{SeMoFlow} (Joint)
& \underline{0.4792}
& 0.0795
& \textbf{0.0250}
& \textbf{0.0317}
& \textbf{0.0020}
& \underline{0.0119} \\

\bottomrule
\end{tabular*}
\end{table*}

We further evaluate SeMoFlow on three Foursquare cities, with results reported in Table~\ref{tab:table2_foursquare}. 
Although 30,000 trajectories are generated for each city and model setting, when fewer than 30,000 real test trajectories are available, metric computation uses a generated subset matched to the size of the corresponding test set.
Owing to the small size of the Foursquare datasets, per-city SeMoFlow underfits, particularly in travel distance, radius of gyration, and recurrence. 
Joint multi-city training substantially mitigates this limitation: the jointly trained model consistently outperforms its per-city counterpart on all six metrics and achieves the lowest (or tied-lowest) travel-distance, gyration, and motif divergences across all three cities, while remaining competitive in POI frequency. 
Recurrence divergence remains higher than that of Traveller, which explicitly models individual travel patterns. 
These results show that sharing mobility knowledge across cities through the unified Semantic ID space is particularly beneficial for data-scarce cities.

\subsection{Motif Structures of Generated Trajectory Data}
\label{sec:Recurrence Patterns}

\begin{figure}[h]
    \centering
    \includegraphics[width=1\linewidth]{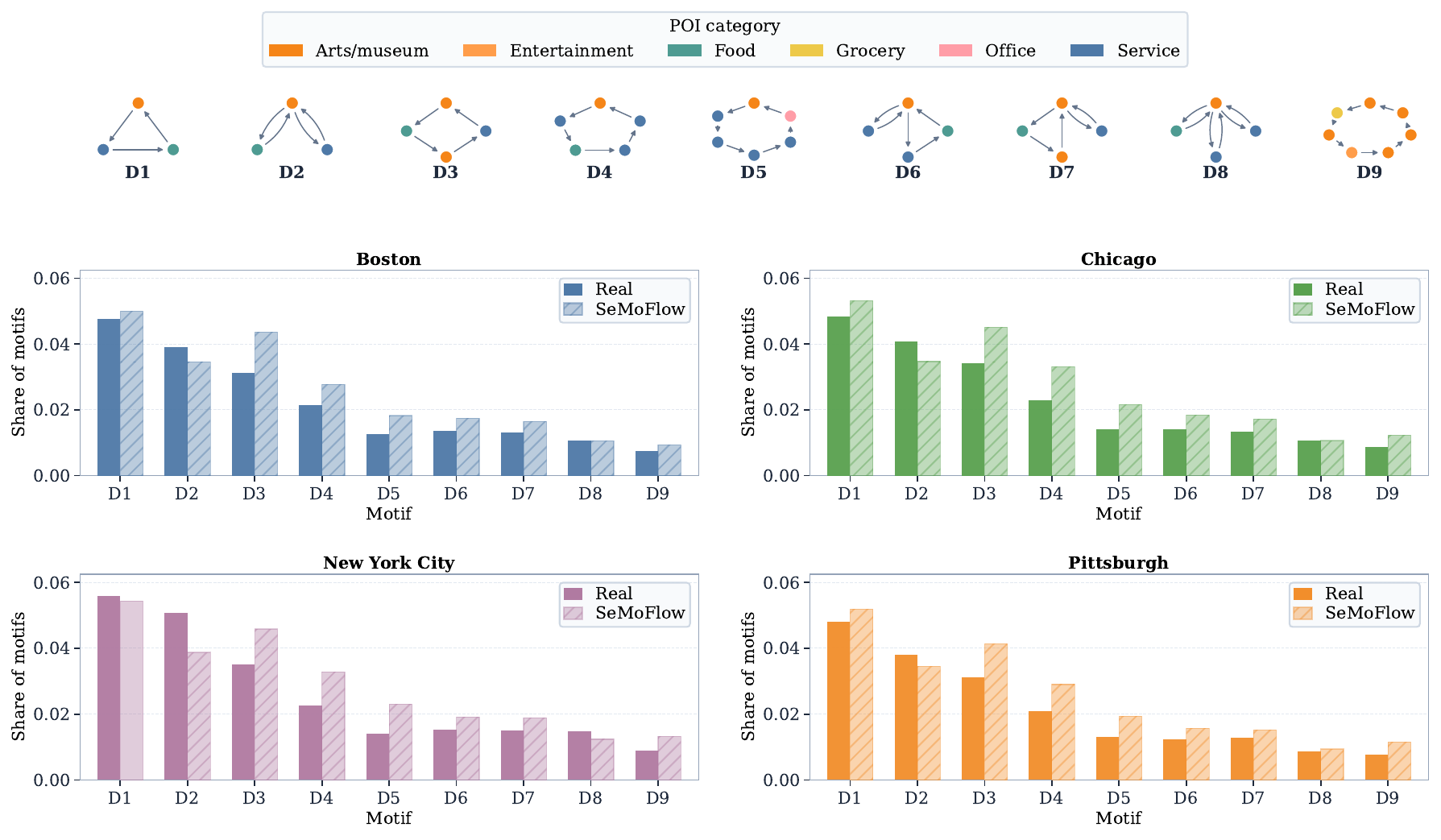}
    \caption{Comparison of daily mobility motifs between real and SeMoFlow-generated trajectories across four cities. The top row illustrates representative motif structures, while the bottom panels compare their frequencies in real and generated trajectories.}
    \label{fig:motif_structures}
\end{figure}

Marginal distributions, such as POI frequency and spatial density, capture where people visit but do not fully describe how visits are organized within a trajectory. 
We therefore evaluate mobility motifs as complementary measures of sequence-level fidelity. Daily motifs summarize the transition structures formed by these visits. Figure~\ref{fig:motif_structures} shows that \textit{SeMoFlow} reproduces the relative frequencies of major daily motifs observed in the real trajectories across all four cities. These results indicate that SeMoFlow captures marginal
visitation statistics while preserving the recurring
sequence structures that characterize human mobility.

%% file: iclr2027_conference.bbl
\begin{thebibliography}{40}
\providecommand{\natexlab}[1]{#1}
\providecommand{\url}[1]{\texttt{#1}}
\expandafter\ifx\csname urlstyle\endcsname\relax
  \providecommand{\doi}[1]{doi: #1}\else
  \providecommand{\doi}{doi: \begingroup \urlstyle{rm}\Url}\fi

\bibitem[Chen et~al.(2025)Chen, Tao, Jiang, Liu, Yu, and Cong]{chen2025enhancing}
Yile Chen, Yicheng Tao, Yue Jiang, Shuai Liu, Han Yu, and Gao Cong.
\newblock Enhancing large language models for mobility analytics with semantic location tokenization.
\newblock In \emph{Proceedings of the 31st ACM SIGKDD Conference on Knowledge Discovery and Data Mining V. 2}, pp.\  262--273, 2025.

\bibitem[Chung et~al.(2014)Chung, Gulcehre, Cho, and Bengio]{chung2014empirical}
Junyoung Chung, Caglar Gulcehre, KyungHyun Cho, and Yoshua Bengio.
\newblock Empirical evaluation of gated recurrent neural networks on sequence modeling.
\newblock \emph{arXiv preprint arXiv:1412.3555}, 2014.

\bibitem[Cuturi(2013)]{cuturi2013sinkhorn}
Marco Cuturi.
\newblock Sinkhorn distances: Lightspeed computation of optimal transport.
\newblock In \emph{Advances in Neural Information Processing Systems (NeurIPS)}, 2013.

\bibitem[Deng et~al.(2025)Deng, Jing, Yang, Qu, Yang, and Cudr{\'e}-Mauroux]{deng2025mirage}
Bangchao Deng, Xin Jing, Tianyue Yang, Bingqing Qu, Dingqi Yang, and Philippe Cudr{\'e}-Mauroux.
\newblock Revisiting synthetic human trajectories: Imitative generation and benchmarks beyond datasaurus.
\newblock In \emph{Proceedings of the 31st ACM SIGKDD Conference on Knowledge Discovery and Data Mining V.1}, pp.\  201--212, 2025.
\newblock \doi{10.1145/3690624.3709180}.
\newblock URL \url{https://doi.org/10.1145/3690624.3709180}.

\bibitem[Feng et~al.(2020)Feng, Yang, Xu, Yu, Wang, and Li]{feng2020movesim}
Jie Feng, Zeyu Yang, Fengli Xu, Haisu Yu, Mudan Wang, and Yong Li.
\newblock Learning to simulate human mobility.
\newblock In \emph{Proceedings of the 26th ACM SIGKDD International Conference on Knowledge Discovery \& Data Mining}, KDD '20, pp.\  3426--3433, New York, NY, USA, 2020. Association for Computing Machinery.
\newblock ISBN 9781450379984.
\newblock \doi{10.1145/3394486.3412862}.
\newblock URL \url{https://doi.org/10.1145/3394486.3412862}.

\bibitem[Gonz{\'a}lez et~al.(2008)Gonz{\'a}lez, Hidalgo, and Barab{\'a}si]{gonzalez2008mobility}
Marta~C. Gonz{\'a}lez, C{\'e}sar~A. Hidalgo, and Albert-L{\'a}szl{\'o} Barab{\'a}si.
\newblock Understanding individual human mobility patterns.
\newblock \emph{Nature}, 453\penalty0 (7196):\penalty0 779--782, 2008.
\newblock \doi{10.1038/nature06958}.
\newblock URL \url{https://doi.org/10.1038/nature06958}.

\bibitem[Guo et~al.(2025)Guo, Hong, Cao, Li, Li, Rong, Prakash, Wang, and Zhao]{guo2025language}
Baoshen Guo, Zhiqing Hong, Lidan Cao, Donghang Li, Junyi Li, Can Rong, Alok Prakash, Shenhao Wang, and Jinhua Zhao.
\newblock Language models meet urban mobility: A data-centric review.
\newblock 2025.

\bibitem[Guo et~al.(2026)Guo, Hong, Li, Wang, and Zhao]{guo2026leveraging}
Baoshen Guo, Zhiqing Hong, Junyi Li, Shenhao Wang, and Jinhua Zhao.
\newblock Leveraging the spatial hierarchy: Coarse-to-fine trajectory generation via cascaded hybrid diffusion.
\newblock In \emph{Proceedings of the 32nd ACM SIGKDD Conference on Knowledge Discovery and Data Mining V. 1}, pp.\  359--370, 2026.

\bibitem[Ho et~al.(2020)Ho, Jain, and Abbeel]{ho2020denoising}
Jonathan Ho, Ajay Jain, and Pieter Abbeel.
\newblock Denoising diffusion probabilistic models.
\newblock \emph{Advances in neural information processing systems}, 33:\penalty0 6840--6851, 2020.

\bibitem[Jiang et~al.(2016)Jiang, Yang, Gupta, Veneziano, Athavale, and Gonz{\'a}lez]{jiang2016}
Shan Jiang, Yingxiang Yang, Siddharth Gupta, Daniele Veneziano, Shounak Athavale, and Marta~C. Gonz{\'a}lez.
\newblock The {TimeGeo} modeling framework for urban mobility without travel surveys.
\newblock \emph{Proceedings of the National Academy of Sciences}, 113\penalty0 (37):\penalty0 E5370--E5378, 2016.
\newblock \doi{10.1073/pnas.1524261113}.
\newblock URL \url{https://doi.org/10.1073/pnas.1524261113}.

\bibitem[Ju et~al.(2025)Ju, Collins, Neves, Kumar, Wang, Zhao, and Shah]{ju2025generative}
Clark~Mingxuan Ju, Liam Collins, Leonardo Neves, Bhuvesh Kumar, Louis~Yufeng Wang, Tong Zhao, and Neil Shah.
\newblock Generative recommendation with semantic ids: a practitioner's handbook.
\newblock In \emph{Proceedings of the 34th ACM International Conference on Information and Knowledge Management}, pp.\  6420--6425, 2025.

\bibitem[Lee et~al.(2022)Lee, Kim, Kim, Cho, and Han]{lee2022autoregressive}
Doyup Lee, Chiheon Kim, Saehoon Kim, Minsu Cho, and Wook-Shin Han.
\newblock Autoregressive image generation using residual quantization.
\newblock In \emph{Proceedings of the IEEE/CVF Conference on Computer Vision and Pattern Recognition (CVPR)}, pp.\  11523--11532, 2022.

\bibitem[Li et~al.(2026)Li, Wang, Zhang, Shi, Koshizuka, Shimizu, and Jiang]{li2026trajflow}
Peiran Li, Jiawei Wang, Haoran Zhang, Xiaodan Shi, Noboru Koshizuka, Chihiro Shimizu, and Renhe Jiang.
\newblock {TrajFlow}: Nation-wide pseudo {GPS} trajectory generation with flow matching models.
\newblock In \emph{International Conference on Learning Representations}, 2026.
\newblock URL \url{https://arxiv.org/abs/2603.15009}.

\bibitem[Li \& He(2026)Li and He]{li2026back}
Tianhong Li and Kaiming He.
\newblock Back to basics: Let denoising generative models denoise.
\newblock In \emph{Proceedings of the IEEE/CVF Conference on Computer Vision and Pattern Recognition}, pp.\  36115--36125, 2026.

\bibitem[Lipman et~al.(2023)Lipman, Chen, Ben-Hamu, Nickel, and Le]{lipman2023flowmatching}
Yaron Lipman, Ricky T.~Q. Chen, Heli Ben-Hamu, Maximilian Nickel, and Matthew Le.
\newblock Flow matching for generative modeling.
\newblock In \emph{International Conference on Learning Representations}, 2023.
\newblock URL \url{https://openreview.net/forum?id=PqvMRDCJT9t}.

\bibitem[Liu et~al.(2026)Liu, Li, Xiao, Li, Min, Tang, and Li]{liu2026allcities}
Bo~Liu, Tong Li, Zhu Xiao, Ruihui Li, Geyong Min, Zhuo Tang, and Kenli Li.
\newblock All cities are equal: A unified human mobility generation model enabled by {LLM}s.
\newblock \emph{arXiv preprint arXiv:2602.19694}, 2026.
\newblock URL \url{https://arxiv.org/abs/2602.19694}.

\bibitem[Liu et~al.(2022)Liu, Gong, and Liu]{liu2022flow}
Xingchao Liu, Chengyue Gong, and Qiang Liu.
\newblock Flow straight and fast: Learning to generate and transfer data with rectified flow.
\newblock \emph{arXiv preprint arXiv:2209.03003}, 2022.

\bibitem[Luo et~al.(2026)Luo, Zhang, Liu, Xu, and Yin]{luo2026traveller}
Yuxiao Luo, Songming Zhang, Kang Liu, Yang Xu, and Ling Yin.
\newblock {Traveller}: Travel-pattern aware trajectory generation via autoregressive diffusion models.
\newblock \emph{Information Fusion}, 127:\penalty0 103766, 2026.
\newblock \doi{10.1016/j.inffus.2025.103766}.
\newblock URL \url{https://doi.org/10.1016/j.inffus.2025.103766}.

\bibitem[Rajput et~al.(2023)Rajput, Mehta, Singh, Keshavan, Vu, Heldt, Hong, Tay, Tran, Samost, Kula, Chi, and Sathiamoorthy]{rajput2023tiger}
Shashank Rajput, Nikhil Mehta, Anima Singh, Raghunandan~Hulikal Keshavan, Trung Vu, Lukasz Heldt, Lichan Hong, Yi~Tay, Vinh~Q. Tran, Jonah Samost, Maciej Kula, Ed~H. Chi, and Maheswaran Sathiamoorthy.
\newblock Recommender systems with generative retrieval.
\newblock In \emph{Advances in Neural Information Processing Systems}, volume~36, 2023.
\newblock \doi{10.52202/075280-0452}.

\bibitem[Schl{\"a}pfer et~al.(2021)Schl{\"a}pfer, Dong, O’Keeffe, Santi, Szell, Salat, Anklesaria, Vazifeh, Ratti, and West]{schlapfer2021universal}
Markus Schl{\"a}pfer, Lei Dong, Kevin O’Keeffe, Paolo Santi, Michael Szell, Hadrien Salat, Samuel Anklesaria, Mohammad Vazifeh, Carlo Ratti, and Geoffrey~B West.
\newblock The universal visitation law of human mobility.
\newblock \emph{Nature}, 593\penalty0 (7860):\penalty0 522--527, 2021.

\bibitem[Schneider et~al.(2013)Schneider, Belik, Couronn{\'e}, Smoreda, and Gonz{\'a}lez]{schneider2013unravelling}
Christian~M Schneider, Vitaly Belik, Thomas Couronn{\'e}, Zbigniew Smoreda, and Marta~C Gonz{\'a}lez.
\newblock Unravelling daily human mobility motifs.
\newblock \emph{Journal of The Royal Society Interface}, 10\penalty0 (84):\penalty0 20130246, 2013.

\bibitem[Siampou et~al.(2026)Siampou, Choudhury, Hsu, Arora, and Shahabi]{siampou2026mepois}
Maria~Despoina Siampou, Shushman Choudhury, Shang-Ling Hsu, Neha Arora, and Cyrus Shahabi.
\newblock Mobility-embedded {POI}s: Learning what a place is and how it is used from human movement.
\newblock In \emph{International Conference on Machine Learning}, 2026.
\newblock URL \url{https://arxiv.org/abs/2601.21149}.

\bibitem[Song et~al.(2024)Song, Ding, Yuan, Liao, and Li]{song2024controllable}
Yiwen Song, Jingtao Ding, Jian Yuan, Qingmin Liao, and Yong Li.
\newblock Controllable human trajectory generation using profile-guided latent diffusion.
\newblock \emph{ACM Transactions on Knowledge Discovery from Data}, 19\penalty0 (1):\penalty0 1--25, 2024.

\bibitem[van~den Oord et~al.(2017)van~den Oord, Vinyals, and Kavukcuoglu]{oord2017neural}
Aaron van~den Oord, Oriol Vinyals, and Koray Kavukcuoglu.
\newblock Neural discrete representation learning.
\newblock In \emph{Advances in Neural Information Processing Systems (NeurIPS)}, 2017.

\bibitem[{Veraset}(2022)]{veraset2022visits}
{Veraset}.
\newblock Visits [dataset].
\newblock Dewey Data, 2022.
\newblock URL \url{https://doi.org/10.82551/TP7E-VA02}.

\bibitem[Wang et~al.(2025)Wang, Huang, Gao, Wang, Huang, and Shang]{wang2025gnprsid}
Dongsheng Wang, Yuxi Huang, Shen Gao, Yifan Wang, Chengrui Huang, and Shuo Shang.
\newblock Generative next {POI} recommendation with semantic {ID}.
\newblock In \emph{Proceedings of the 31st ACM SIGKDD Conference on Knowledge Discovery and Data Mining}, pp.\  2904--2914, 2025.
\newblock \doi{10.1145/3711896.3736981}.

\bibitem[Wang et~al.(2024{\natexlab{a}})Wang, Jiang, Yang, Wu, Onizuka, Shibasaki, Koshizuka, and Xiao]{wang2024llmob}
Jiawei Wang, Renhe Jiang, Chuang Yang, Zengqing Wu, Makoto Onizuka, Ryosuke Shibasaki, Noboru Koshizuka, and Chuan Xiao.
\newblock Large language models as urban residents: An {LLM} agent framework for personal mobility generation.
\newblock In \emph{Advances in Neural Information Processing Systems}, volume~37, pp.\  124547--124574, 2024{\natexlab{a}}.
\newblock \doi{10.52202/079017-3957}.
\newblock URL \url{https://proceedings.neurips.cc/paper_files/paper/2024/hash/e142fd2b70f10db2543c64bca1417de8-Abstract-Conference.html}.

\bibitem[Wang et~al.(2024{\natexlab{b}})Wang, Zheng, Liang, Liu, and Song]{wang2024cola}
Yu~Wang, Tongya Zheng, Yuxuan Liang, Shunyu Liu, and Mingli Song.
\newblock {COLA}: Cross-city mobility transformer for human trajectory simulation.
\newblock In \emph{Proceedings of the ACM Web Conference 2024}, pp.\  3509--3520, 2024{\natexlab{b}}.
\newblock \doi{10.1145/3589334.3645469}.
\newblock URL \url{https://doi.org/10.1145/3589334.3645469}.

\bibitem[Wang et~al.(2026)Wang, Yang, Wang, Xu, Xu, Li, Xiao, and Jiang]{wang2026ellmob}
Yusong Wang, Chuang Yang, Jiawei Wang, Xiaohang Xu, Jiayi Xu, Dongyuan Li, Chuan Xiao, and Renhe Jiang.
\newblock {ELLMob}: Event-driven human mobility generation with self-aligned {LLM} framework.
\newblock In \emph{International Conference on Learning Representations}, 2026.
\newblock URL \url{https://openreview.net/forum?id=MPYsaBgZIT}.

\bibitem[Wei et~al.(2025{\natexlab{a}})Wei, Ning, Chen, Qiu, Hou, Xie, Yang, Hua, and He]{wei2025cofirec}
Tianxin Wei, Xuying Ning, Xuxing Chen, Ruizhong Qiu, Yupeng Hou, Yan Xie, Shuang Yang, Zhigang Hua, and Jingrui He.
\newblock Cofirec: Coarse-to-fine tokenization for generative recommendation, 2025{\natexlab{a}}.
\newblock URL \url{https://arxiv.org/abs/2511.22707}.

\bibitem[Wei et~al.(2024)Wei, Lin, Guo, Lin, Huang, Xiang, Bai, and Wan]{wei2024diff}
Tonglong Wei, Youfang Lin, Shengnan Guo, Yan Lin, Yiheng Huang, Chenyang Xiang, Yuqing Bai, and Huaiyu Wan.
\newblock Diff-rntraj: A structure-aware diffusion model for road network-constrained trajectory generation.
\newblock \emph{IEEE Transactions on Knowledge and Data Engineering}, 2024.

\bibitem[Wei et~al.(2025{\natexlab{b}})Wei, Lin, Zhou, Wen, Hu, Guo, Lin, Cong, and Wan]{wei2025transfertraj}
Tonglong Wei, Yan Lin, Zeyu Zhou, Haomin Wen, Jilin Hu, Shengnan Guo, Youfang Lin, Gao Cong, and Huaiyu Wan.
\newblock {TransferTraj}: A vehicle trajectory learning model for region and task transferability.
\newblock In \emph{Advances in Neural Information Processing Systems}, volume~38, pp.\  109941--109968, 2025{\natexlab{b}}.
\newblock \doi{10.52202/085713-3674}.
\newblock URL \url{https://proceedings.neurips.cc/paper_files/paper/2025/hash/9f02b7c937447b6857b21d31d5c344e6-Abstract-Conference.html}.

\bibitem[Wongso et~al.(2026)Wongso, Li, Prabowo, Lin, Chen, Xue, and Salim]{wongso2026trajdlm}
Wilson Wongso, Lihuan Li, Arian Prabowo, Xiachong Lin, Baiyu Chen, Hao Xue, and Flora~D Salim.
\newblock Trajdlm: Topology-aware block diffusion language model for trajectory generation.
\newblock \emph{arXiv preprint arXiv:2605.10020}, 2026.

\bibitem[Xu et~al.(2026{\natexlab{a}})Xu, Cai, Jiang, Hong, Tian, and Wang]{xu2026geogen}
Rongchao Xu, Kunlin Cai, Lin Jiang, Zhiqing Hong, Yuan Tian, and Guang Wang.
\newblock {GeoGen}: A two-stage coarse-to-fine framework for fine-grained synthetic location-based social network trajectory generation.
\newblock \emph{Proceedings of the AAAI Conference on Artificial Intelligence}, 40\penalty0 (2):\penalty0 1373--1381, 2026{\natexlab{a}}.
\newblock \doi{10.1609/aaai.v40i2.37111}.
\newblock URL \url{https://ojs.aaai.org/index.php/AAAI/article/view/37111}.

\bibitem[Xu et~al.(2026{\natexlab{b}})Xu, Jiang, Yu, Li, Liu, Zhang, Tian, and Wang]{xu2026mobidiff}
Rongchao Xu, Lin Jiang, Dahai Yu, Ximiao Li, Taichi Liu, Desheng Zhang, Yuan Tian, and Guang Wang.
\newblock {MobiDiff}: Semantic-aware multi-channel discrete diffusion for human mobility data generation.
\newblock \emph{arXiv preprint arXiv:2607.08357}, 2026{\natexlab{b}}.
\newblock URL \url{https://arxiv.org/abs/2607.08357}.

\bibitem[Xu et~al.(2026{\natexlab{c}})Xu, Jiang, Yu, Li, and Wang]{xu2026synhat}
Rongchao Xu, Lin Jiang, Dahai Yu, Ximiao Li, and Guang Wang.
\newblock {SynHAT}: A two-stage coarse-to-fine diffusion framework for synthesizing human activity traces.
\newblock \emph{Proceedings of the ACM on Interactive, Mobile, Wearable and Ubiquitous Technologies}, 10\penalty0 (2):\penalty0 1--35, 2026{\natexlab{c}}.
\newblock \doi{10.1145/3810213}.
\newblock URL \url{https://doi.org/10.1145/3810213}.

\bibitem[Yabe et~al.(2024)Yabe, Tsubouchi, Shimizu, Sekimoto, Sezaki, Moro, and Pentland]{yabe2024yjmob100k}
Takahiro Yabe, Kota Tsubouchi, Toru Shimizu, Yoshihide Sekimoto, Kaoru Sezaki, Esteban Moro, and Alex Pentland.
\newblock Yjmob100k: City-scale and longitudinal dataset of anonymized human mobility trajectories.
\newblock \emph{Scientific Data}, 11\penalty0 (1):\penalty0 397, 2024.

\bibitem[Yang et~al.(2016)Yang, Zhang, and Qu]{yang2016participatory}
Dingqi Yang, Daqing Zhang, and Bingqing Qu.
\newblock Participatory cultural mapping based on collective behavior data in location-based social networks.
\newblock \emph{ACM Transactions on Intelligent Systems and Technology (TIST)}, 7\penalty0 (3):\penalty0 1--23, 2016.

\bibitem[Zhu et~al.(2023)Zhu, Ye, Zhang, Zhao, and Yu]{zhu2023difftraj}
Yuanshao Zhu, Yongchao Ye, Shiyao Zhang, Xiangyu Zhao, and James Yu.
\newblock Difftraj: Generating gps trajectory with diffusion probabilistic model.
\newblock \emph{Advances in Neural Information Processing Systems}, 36:\penalty0 65168--65188, 2023.

\bibitem[Zhu et~al.(2024)Zhu, Yu, Zhao, Liu, Ye, Chen, Zhang, Wei, and Liang]{controltraj2024}
Yuanshao Zhu, James J.~Q. Yu, Xiangyu Zhao, Qidong Liu, Yongchao Ye, Wei Chen, Zijian Zhang, Xuetao Wei, and Yuxuan Liang.
\newblock {ControlTraj}: Controllable trajectory generation with topology-constrained diffusion model.
\newblock In \emph{Proceedings of the 30th ACM SIGKDD Conference on Knowledge Discovery and Data Mining}, pp.\  4676--4687, 2024.
\newblock \doi{10.1145/3637528.3671866}.

\end{thebibliography}
